\documentclass[article,amsmath,amssymb,superscriptaddress,nofootinbib,onecolumn]{revtex4-1}

\usepackage{graphicx}
\usepackage{amsmath,pgfmath,pgffor}
\usepackage[usenames, dvipsnames]{color}
\usepackage{indentfirst}
\usepackage{hyperref}

\usepackage[T1]{fontenc}
\usepackage[english, american]{babel}
\usepackage{amsmath}
\usepackage[utf8]{inputenc}

\usepackage{makeidx}
\usepackage{bigints}

\usepackage{colortbl}

\usepackage{feyn}

\usepackage{color}

\hypersetup{%
    pdfborder = {0 0 0},
    colorlinks,
    citecolor=magenta,
    linkcolor=blue,
}

\begin{document}

\title{Dynamical Mean-Field Theory and Aging Dynamics}
\author{Ada Altieri} 
\affiliation{Laboratoire de Physique de l'\'Ecole Normale Supérieure, Université PSL, CNRS, Sorbonne Université, Université Paris-Diderot, Paris, France}
\author{Giulio Biroli}
\affiliation{Laboratoire de Physique de l'\'Ecole Normale Supérieure, Université PSL, CNRS, Sorbonne Université, Université Paris-Diderot, Paris, France}
\author{Chiara Cammarota}
\affiliation{Department of Mathematics, King's College London, Strand London WC2R 2LS, UK}

\begin{abstract}
Dynamical Mean-Field Theory (DMFT) replaces the many-body dynamical problem with one for a single degree of freedom in a thermal bath whose features are determined self-consistently. By focusing on models with soft disordered $p$-spin interactions, we show how to incorporate the mean-field theory of aging within dynamical mean-field theory. We study cases with only one slow time-scale, corresponding statically to the one-step replica symmetry breaking (1RSB) phase, and cases with an infinite number of slow time-scales,  
corresponding statically to the full replica symmetry breaking (FRSB) phase. For the former, we show that the effective temperature of the slow degrees of freedom is fixed by requiring critical dynamical behavior on short time-scales, {\it i.e.} marginality. For the latter, we find that aging on an infinite number of slow time-scales is governed by a stochastic equation where the clock for dynamical evolution is fixed by the change of effective temperature, hence obtaining a dynamical derivation of the stochastic equation at the basis of the FRSB phase. 
Our results extend the realm 
of the mean-field theory of aging to all situations where DMFT holds. 

\end{abstract}

\maketitle

\tableofcontents
\newpage

\section{Introduction}
Many metastable states, slow dynamics, and aging are hallmarks of glassiness.
The study of mean-field models has been instrumental in revealing these features and understanding such phenomena. The first analysis of the dynamics of mean-field glassy systems was pioneered by Sompolinsky and Zippelius \cite{Sompolinsky1982}, who were the first to obtain dynamical mean-field equations for glassy systems. At that time, the interest was mainly on the {\it equilibrium} properties. Later, the focus shifted on {\it off-equilibrium}, and an exact analysis of the aging dynamics was worked out for a disparate set of models \cite{Cugliandolo1993,Cugliandolo1994_aging,Franz1994,Cugliandolo1996RM}. The peculiarity of these models is that their dynamical mean-field equations
simplify considerably. In fact, instead of dealing with a self-consistent stochastic process, representing the dynamics of a single degree of freedom in the self-consistent bath formed by the rest of the system, their dynamics can be studied via a closed set of integro-differential equations on correlation and response functions, a fact that played an important role in their exact analysis. 

The picture resulting from these works goes beyond the exact solution of these simplified models and provides a general scenario for aging dynamics 
for all mean-field glassy systems (see \cite{Folena2019,Bernaschi2019} for very recent surprises). 
Yet, a complete dynamical mean-field theory of aging that applies to generic cases where the dynamics can be studied only through the analysis of the self-consistent stochastic process is still missing. This is not a mere technical curiosity, it is actually relevant for the study of topics as diverse as ecosystem dynamics, the glass transition, and optimization dynamics of neural networks  \cite{RoyDMFT,Agoritsas2018, Agoritsas2019out}.

The aim of our work is to extend the mean-field theory of aging to generic dynamical mean-field theories (DMFT). We take the mean-field picture of aging \cite{Cugliandolo2002LH, Cugliandolo2003slow} as a starting point, and work out its main 
implications for DMFT. We make use of many results obtained along the years. 
In particular, we combine the ideas put forward by Sompolinsky and Zippelius \cite{sompolinsky1981time,Sompolinsky1982} on dynamics on very large time-scales with the ones developed by Cugliandolo and Kurchan on effective temperatures and slow thermal baths \cite{cugliandolo2000}. 

We shall show how to obtain explicit equations on the correlation and the response of the systems on diverging time-scales. In particular, in cases (called $1$RSB-like) where the slow dynamics is 
described by only one diverging time-scale we find that the effective temperature of the slow degrees of freedom is determined by the condition that the dynamics on fast time scales is marginal, {\it i.e.} the correlation function decreases as a power law in time. 
In cases where the slow dynamics is 
described by an infinite set of diverging time-scales (so-called Full RSB-like) we find that the slow dynamics contribution is given by a stochastic equation where the clock for dynamical evolution is fixed by the change of the effective temperature. These results generalise the ones found in simplified models. The latter one provides a dynamical derivation of the stochastic equation at the basis of full-replica symmetry breaking \cite{MPV}. 

We will comment in the Conclusions on possible extensions, and applications of our results to theoretical ecology \cite{RoyDMFT}, out-of-equilibrium dynamics of hard spheres in the limit of infinite dimensions \cite{Agoritsas2019out, Agoritsas2019}, and gradient-descent based algorithms for non-convex optimization problems \cite{Sarao,Sarao2020}.

The paper is organised as follows: in Section \ref{DMFT_eqs} we will present a class of disordered models defined by $p$-spin interactions for which the DMFT formalism applies; in Section \ref{1timescale} the \emph{aging} hypothesis is described in full generality along with the discussion on two different kinds of dynamically broken phases, according to a $1$RSB and a Full RSB Ansatz respectively. 
We will then present our formalism based on a sharp separation of time scales, focusing first on the (fast) TTI regime in Sec.(\ref{TTI}) and then on the slow dynamical phase corresponding to aging, in Sec. (\ref{aging_1RSB}). 
Next Sec. (\ref{Dyson}) will be then devoted to the definition of the effective temperature for models that display a one-step replica symmetry breaking solution: the key result relies on the determination of the marginal stability condition for two different classes of $p$-spin models, with continuous and discrete variables respectively.
In Section \ref{full} we will extend our predictions to a pairwise interaction model, namely the classical Sherrington-Kirkpatrick model. We will prove that also in this case we can write a dynamical effective stochastic process for the effective fields, which exactly maps into the equation for ultrametricity as it was obtained in the past in a static formalism.
Finally, in Section \ref{conclusions} we will present our conclusive remarks and some perspectives for future investigations in related fields.

\section{Dynamical mean-field equations}
\label{DMFT_eqs}

\subsection{Models with $p$-spin interactions}
\label{SGmodels}
In order to develop the theoretical framework we focus on a simple class of mean-field models, but our results can be generalized to more complex cases. 
The elementary degrees of freedom of the models are real variables, that we shall call spins and denote as $s_i$ ($i=1,\dots,N$). Each spin is subjected to an external potential $V(s_i)$.  
The interaction part of the Hamiltonian is given by random $p$-spin interactions: 
\begin{equation}
H_I=-\sum \limits_{i_1< ...<i_p} J_{i_1...i_p} s_{i_1}...s_{i_p} 
\label{H_Ising_p}
\end{equation}
where the $J_{i_1...i_p}$s are quenched random variables distributed according to the law 
\begin{equation}
P(J_{i_1...i_p})= \sqrt{\frac{N^{p-1}}{ \pi p!}} \exp{\left(-\frac{J^2_{i_1...i_p}N^{p-1}}{p!} \right)} \ . 
\end{equation}
The scaling in $N$ is chosen in such a way to have a well-defined limit as $N \rightarrow \infty$.
The Hamiltonian of the system is therefore
\begin{equation}
H=H_I+\sum_i V(s_i) \ .    
\end{equation}
By tuning the potential $V(s_i)$ one can recover standard $p$-spin models with spherical spins, which  
assume continuous values $s_i\in \mathbb{R}$, or Ising 
spins, which assume discrete values $s_i \in {\pm 1}$. In the first case the addition of a soft spherical constraint is implemented by choosing $V(s_i)=0.5\lambda s_i^2$ and gives origin to the class of spherical $p$-spin models, which for $p\geq 3$ provide a mean-field paradigm for structural glasses \cite{kirkpatrick1987p, Thirumalai1988, Kirkpatrick1989, Kirkpatrick1989random, Castellani2005}. 
To obtain hard-spins, one can choose $V(s_i)=\alpha (s_i^2-1)^2$ and take the limit $\alpha \rightarrow \infty$. This leads to the so-called Ising $p$-spin models and, for $p=2$, to the Sherrington-Kirkpatrick (SK) model of spin glasses  \cite{Sherrington1975, Kirkpatrick1978}.

\subsection{Dynamical Mean-Field Theory formalism}
The dynamics we are going to focus on is induced by Langevin equations that read:
\begin{equation}
\frac{d s_i(t)}{dt}=-\frac{\partial V}{\partial s_i} +\frac{1}{(p-1)!} \sum \limits_{i_2...i_p} J_{i,i_2...i_p} s_{i_2}...s_{i_p} +\eta_i(t) \ .
\end{equation}
The first and second term appearing on the right hand side (RHS) correspond to the derivative 
of the Hamiltonian with respect to the given spin $s_{i_1}$, whereas the noise is (for simplicity) $\delta$-correlated,  $\langle \eta_i(t)\eta_j(t')\rangle = 2 T \delta_{ij} \delta(t-t')$ ($T$ is the temperature).
We shall consider a high-temperature like initial condition at $t=0$ given by a non-interacting product measure on the spins:  $P({s_i},t=0)=\prod_{i=1}^NP_0(s_i)$. 
One can then write the corresponding generating functional in terms of a bare contribution and a $J$-dependent term, which has to be eventually averaged over the disorder \cite{Sompolinsky1982,kirkpatrick1987p, Crisanti1993}, and from it obtain the dynamical mean-field equations. Alternatively, one can use the dynamical cavity method \cite{MPV,RoyDMFT}. These derivations are standard. Hence, we directly state the final result, \textit{i.e.} the DMFT equation that reads:
\begin{equation}
\dot{s}(t)= -\frac{\partial V(s(t))}{\partial s}+ \frac{p(p-1)}{2} \int_0^t dt'' R(t,t'') C^{p-2}(t,t'')s(t'')+\xi(t) \ ,
\label{dot-s}
\end{equation}
where the noise is such that 
\begin{equation}
\langle \xi(t) \xi(t') \rangle = 2T\delta(t-t')+\frac{p}{2} C^{p-1}(t,t') \  .
\label{noise-noise}
\end{equation}
The first contribution corresponds to the usual noise, whereas the second one accounts for the interaction with the rest of the system. 
The correlation and the response functions, $C(t,t')$ and $R(t,t')$, are defined respectively as
\begin{equation}
\begin{split}
& C(t,t')= \frac{1}{N} \sum_i s_i(t) s_i(t')  \ ,\\
& R(t,t')=  \frac{1}{N} \sum_i \left . \frac{\delta s_i(t)}{\delta h_i(t')}\right \vert_{h_i=0} \ ,
\end{split}
\end{equation}
where $h_i$ is an external field linearly coupled to $s_i$.
As $N \rightarrow \infty$, these quantities converge to a non-fluctuating value. 
The DMFT equation has to be solved self-consistently, \textit{i.e.} one has to find $C(t,t')$ and $R(t,t')$ such that the stochastic process in Eq. (\ref{dot-s}), with initial condition given by $P_0(s)$, leads to correlation and response functions equal to $C(t,t')$ and $R(t,t')$\footnote{It is possible to show that there is a unique solution respecting causality.}.
In very specific instances, \textit{e.g.} in the so-called spherical limit, the problem simplifies and $C(t,t')$ and $R(t,t')$ can be shown to satisfy closed-form integro-differential equations. Note that those closed equations have formally the same structure as Mode-Coupling Theory equations for structural glasses \cite{Bengtzelius}.

In general, this does not happen and one has to deal with the self-consistent process defined above.  
The aim of our work is to show how its solution can be handled for aging dynamics.

\section{Slow Time-Scales: Aging and dynamical phases}
\label{1timescale}

As we discussed in the Introduction, one key feature of glassy systems is that they display slow and aging dynamics after a quench from high to low temperature. 
The behavior at a large time $t_w$ after the quench is characterized by: (i) power law (or even slower) relaxation of one-time quantities, and (ii) a decorrelation time that grows with the time $t_w$.\\
The theoretical analysis based on mean-field models performed in the 90s has unveiled that there are at least two different 
classes of aging dynamics, correspondingly to two 
different classes of free-energy landscapes \cite{Cugliandolo2003slow}. We now recall their main salient features.

\subsection{Two classes of landscapes}

In the case of mean-field glassy models one can give a precise meaning to the free-energy landscape, which is obtained from the TAP free-energy \cite{TAP}. The number of minima, and more generally of critical points of given index, has been computed and analysed thoroughly \cite{Cavagna1998,Fyodorov2004, Fyodorov2007, Bray2007}, even rigorously in recent years \cite{Auffinger2011, Auffinger2013}. These works established the existence of two very different classes of landscapes, which are related to different thermodynamical and dynamical properties:   
\begin{itemize}
\item \textbf{Spin-glass landscapes}: In this case, the number of free-energy minima is not exponential in the system size. The free-energy barriers are expected to be sub-extensive, and despite quenches to low temperature induce aging, one-time observables converge to their equilibrium thermodynamic limit. For instance, the asymptotic value of the  energy density coincides\footnote{Note that we always consider the case in which the thermodynamic limit is taken from the start, \textit{i.e.} asymptotic values corresponds to the specific limit order $ \lim \limits_{t, t'\rightarrow \infty}  \lim \limits_{N \rightarrow \infty} $.} with the equilibrium value obtained within the static approach, $E_\infty=E_\text{eq}$. Moreover, there is a strong connection between the asymptotic aging dynamics and the thermodynamics \cite{Franz1998}. In fact, the system asymptotically approaches the marginal free-energy minima that are relevant for equilibrium properties. Finally, the dynamical transition at which the system falls out of equilibrium takes place at the same critical temperature of the thermodynamic spin-glass transition. 
%the dynamical transition coincides with the static transition temperature, the latter corresponding to the appearance of a non-zero overlap parameter and the inapplicability of the RS solution, due to its incorrectness in the low-temperature regime. More interestingly, it one considers single-time intensive quantities, as the energy density, in this specific limit order $ \lim \limits_{t, t'\rightarrow \infty}  \lim \limits_{N \rightarrow \infty} $, the asymptotic energy coincides with the equilibrium value obtained within the static approach, $E_\infty=E_\text{eq}$. 
\item \textbf{Simple structural glass landscapes}: In this case, the number of free-energy minima is exponential in the system size. The free-energy barriers are extensive, and one-time observables do not converge to their equilibrium value. A connection with free-energy minima still holds. In fact, long-time aging dynamics approaches free-energy minima with the largest basin of attraction, which are generally not the ones relevant for equilibrium thermodynamic properties. Starting from random initial conditions, \text{i.e.} quenches from infinite temperature, these have been identified as the typical most numerous minima that are marginally stable (called \emph{threshold states}). In this case, the dynamical transition at which the system falls out of equilibrium takes place at higher temperature, $T_d$, than the thermodynamic glass transition $T_s$.

\end{itemize}

As we have recalled above, marginally stable free-energy minima play a key role in aging dynamics. Marginality means that the free-energy Hessian matrix at the minima is characterised by arbitrary small eigenvalues.
There are models characterised by more complicated free-energy landscapes that combine features of spin-glass and structural glass landscapes, as for instance when a Gardner phase transition takes place (a likely general feature for structural glass models at low enough temperature) \cite{Gardner1985, Charbonneau2014, Rizzo2013}. These cases are left for future works, and hence not described in detail here. 

The relationship recalled above between free-energy minima and aging dynamics is at the core of the \emph{weak ergodicity breaking} hypothesis, that has been proven to hold in a large class of systems, in particular the instances we are going to discuss in the rest of the paper. Recent results showed that the general situation can be more intricate: in particular starting the quench from finite (and not infinite) temperature it was shown that for mixed $p$-spin models \cite{Folena2019} the long-time dynamics approach marginally stable free-energy minima that are not the \emph{threshold} ones. In numerical simulations in spin-glass models analysed on sparse graphs \cite{Bernaschi2019},  the authors claimed that the weak ergodicity breaking does not hold.

\subsection{Aging and its two dynamical regimes}
\label{Aging_intro}

In order to analyse aging dynamics, in particular within mean-field, it is useful to make extensive use of long-time limit analysis, which allows for a sharp timescale separation between a fast regime, in which rapid degrees of freedom equilibrate, and a slow regime displaying violation of fluctuation-dissipation relations and non-equilibrium phenomena \cite{Bouchaud1998, Cugliandolo2003slow}. The existence of two-time sectors have been explicitly shown to hold for certain class of mean-field models \cite{Cugliandolo1993, Monthus1996, Cugliandolo2008, Franz2000}. %It has ??
The exploitation of such time separation stands at the core of mean-field analysis of aging dynamics \cite{Cugliandolo1993, Cugliandolo1994_aging, Cugliandolo1994off, Bouchaud1998, Cugliandolo1998_quantum, Bouchaud2000_soft, Cugliandolo2003slow, Franz1994, Biroli2005, Cugliandolo2008}. Cutting a long story short, we directly recall the form the correlation function takes in the long time limit $t, t'\rightarrow \infty$:
\begin{equation}
C(t,t')=C_{\text{TTI}}(t-t')+ C_{\text{A}}(t,t') \ .
\label{C_2sectors}
\end{equation}
In the time-translation invariant (TTI) sector, which corresponds to $t$, $t' \gg 1$ with $(t-t')$ of order one, only the first term on the RHS gives a non vanishing contribution and accordingly
\begin{equation}
    C_{\text{TTI}}(0)=q_d-q_1 \ , \hspace{0.9cm}  C_{\text{TTI}}(\infty)=0 \ .
\end{equation}
The overlap $q_d$ is by definition equal to $C(t,t)$, whereas $q_1$ corresponds to the plateau value of the correlation function separating TTI and aging regime, see Fig. \ref{Correlation}. \\
The aging sector corresponds to $t$, $t' \gg 1$, and $(t-t')$ which diverges together with $t,t'$. In this regime $C_{\text{TTI}}$ is zero and the only contribution to $C(t,t')$ is given by $C_\text{A}(t,t')$, which satisfies the boundary condition:
\begin{equation}
C_\text{A}(t,t)=q_1 \ .    
\end{equation}
The response function displays an analogous behavior which can be decomposed in a TTI and an aging contribution. In the TTI sector the response function verifies the fluctuation-dissipation theorem, \textit{i.e.} 
$R_\text{TTI}(\tau)=-\frac{1}{T} \frac{ d C_\text{TTI}(\tau)}{ d \tau}$. This is natural since 
degrees of freedom contributing to the TTI regime relax on a finite time-scale and, hence, equilibrate at long times. \\
The behavior in the aging regime depends on the dynamical phase, in particular whether there is only one or multiple diverging time-scales. 

\begin{itemize}

\item \textbf{One diverging time-scale: the $\mathbf{1}$RSB dynamical ansatz}
In the simplest scenario, the aging regime 
is described by a single diverging timescale. The corresponding dynamical Ansatz reads:
\begin{equation}
C_\text{A}(t,t')= \mathcal{C}\left[ \frac{\widehat{h}(t')}{\widehat{h}(t)} \right]  \ , \hspace{0.5cm} R_\text{A}(t,t')= \dot{\widehat{h}}(t') \mathcal{R}\left[ \frac{\widehat{h}(t')}{\widehat{h}(t)} \right]
\label{reparam}
\end{equation}
where $\widehat{h}(t)$ is a monotonously increasing function that corresponds to the relaxation timescale of the slow degrees of freedom (from now on we consider $t>t'$). It depends on $t$ because the system is aging: the older is the system, the slower is the relaxation.   
An important and highly non-trivial relationship between correlation and response holds in this regime: $R_\text{A}(t,t')=\frac{x}{T}\partial_{t'}C_{\text{A}}(t,t')$. This is a generalization of the fluctuation-dissipation relation for the aging regime (with an effective temperature 
$T_{eff}$ defined by $x=T/T_\text{eff}$).
This aging behavior has been found in models characterized by simple structural glass 
landscapes, and it is the dynamical counterpart of the $1$RSB static Ansatz.

\begin{figure}[h]
\center
\includegraphics[scale=0.67]{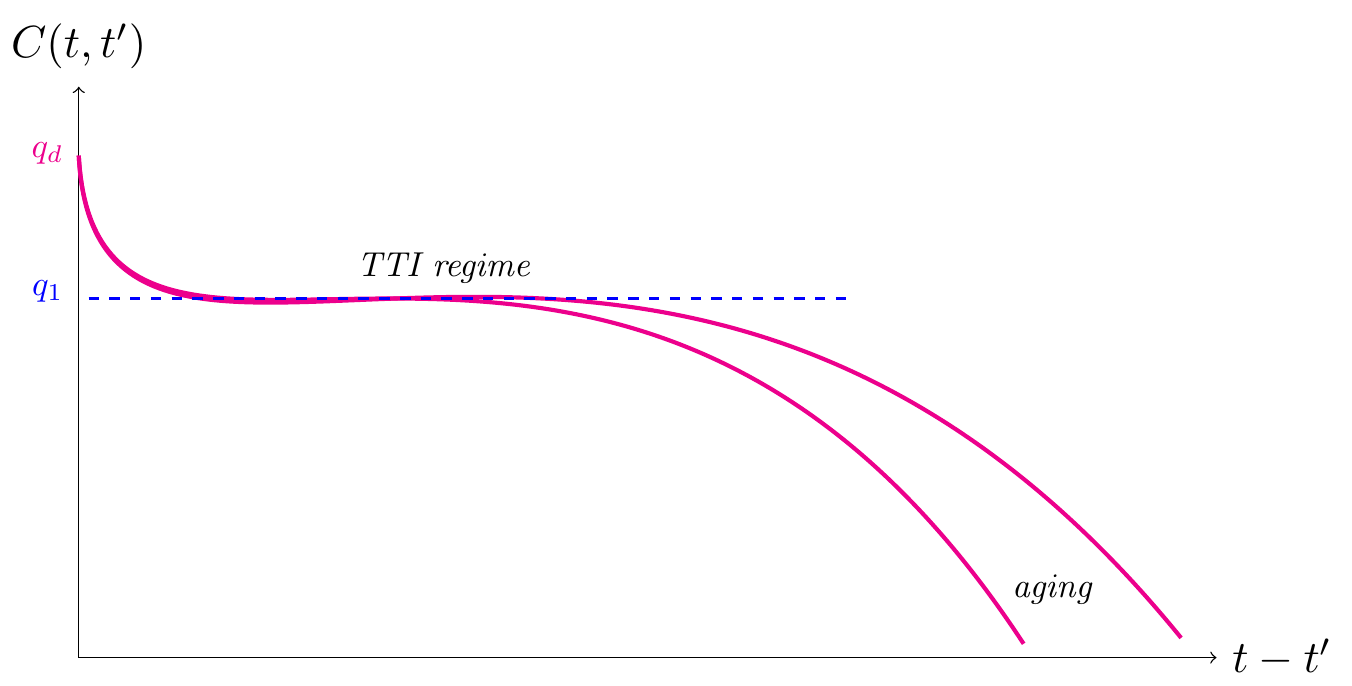}
\caption{The correlation function $C(t,t')$, which depends generically on two times scales, displays a decreasing trend from the maximum value $q_d$ to the plateau value whose height coincides with $q_1$ and signals the onset of non-ergodicity. The escape from the plateau is regulated by the function $C_{\text{A}}(t,t')$.}
\label{Correlation}
\end{figure}

\item \textbf{Infinitely many diverging time-scale: the Full RSB dynamical Ansatz}

This case  is characterised by infinitely many diverging timescales. 
A monotonously increasing function $\widehat{h}_i(t)$ is associated to each timescale $i$. The aging contribution to the correlation function can be written as a combination of rescaled functions $\mathcal{C}_i$ associated to each timescale  \cite{Bouchaud1998}:
\begin{equation}
C_{\text{A}}(t, t')=\sum_i \mathcal{C}_i \left[ \frac{\widehat{h}_i(t')}{\widehat{h}_i(t)} \right] 
\end{equation}
where $\mathcal{C}_i(1)$ is equal to $q_i-q_{i-1}$ and $\mathcal{C}_i(0)=0$ (remember that $t>t'$). Each $\mathcal{C}_i$ describes the drop of the correlation from $q_i$ to $q_{i-1}$ that takes place within the timescale $i$.   
A generalized fluctuation-dissipation relation is valid within each time-scale, \textit{i.e.} 
$R_\text{A}(t,t')=\frac{x_i}{T}\partial_{t'}C_{\text{A}}(t,t')$ for $t,t'$ such that $0< \frac{\widehat{h}_i(t')}{\widehat{h}_i(t)}< 1$. 
We have described the aging Ansatz in terms of a discrete set of timescales. One can take the limit of an infinite number of timescale assuming that that all the differences  $q_i-q_{i-1}$ goes to zero and at the same time the number of timescales goes to infinity. In this case, it is useful to introduce the function $x(q)$ which relates $x_i$ with $q_i$.  
The aging behaviour just described is the dynamical counterpart of the Full RSB static Ansatz.
\end{itemize}

\subsection{Fast and slow noises}
Within DMFT the single spin stochastic equation is characterised by an effective noise that take into account both the thermal noise and the interaction with the rest of the system. Given that the system is characterised by several timescales, so does the noise. To make explicit these different contributions, we express  $\xi(t)$ as the sum of two independent Gaussian noise contributions, $\xi_{\text{TTI}}(t)$ and $\xi_{A}(t)$, such that  
\begin{equation}\label{noise-noiseA}
\langle \xi_{\text{TTI}}(t) \xi_{\text{TTI}}(t') \rangle = 2T\delta(t-t')+{{\frac{p}{2}\left[ C_\text{TTI}(t-t')+q_1 \right]^{p-1} -\frac{p}{2}q_1^{p-1}}}
\  ,
\end{equation}
\begin{equation}\label{noise-noiseB}
\langle \xi_{\text{A}}(t) \xi_{\text{A}}(t') \rangle={{\frac{p}{2}C_{\text{A}}(t,t')^{p-1}}} \ .
\end{equation}

By choosing the covariances in this way, the sum  $\xi(t)=\xi_{\text{TTI}}(t)+\xi_{\text{A}}(t)$ 
leads to a correct representation of the noise in the asymptotic limit $t,t'\gg 1$. The slow noise $\xi_{A}(t)$ can be further decomposed in multiple contributions if there are many slow timescales. Using the notation introduced above for the Full RSB dynamical Ansatz, one can introduce independent Gaussian noise contributions $\xi_{A,i}(t)$ for each slow timescale. In order to have a correct representation of the noise in the asymptotic limit, the covariance of the $\xi_{A,i}(t)$s has to be chosen in the following way:
\begin{equation}
    \langle \xi_{\text{A,i}}(t) \xi_{\text{A,i}}(t')\rangle={{\frac{p}{2} \left(\mathcal{C}_i \left[ \frac{\widehat{h}_i(t')}{\widehat{h}_i(t)} \right]
 +q_{i-1}\right)^{p-1}-\frac{p}{2}q_{i-1}^{p-1} }}\,.
\end{equation}

\section{Fast and Slow Time-Scales: Analysis of the TTI and Aging Regimes}
In the following we show how to disentangle the regimes of fast and slow time-scales in the analysis. As we shall show, in the first regime the system is in quasi-equilibrium and one can study its corresponding quasi-equilibrium dynamics. In the second regime, instead, the system is 
evolving very slowly (the slower the older is the system). This leads to an almost adiabatic change of some of the parameters determining the fast dynamics. 
We will obtain the probability distribution of such parameters along the aging dynamics. All that will allow us to find all the quantities of interest to characterise aging dynamics, except the effective temperature to which we come back in the next two sections.
\subsection{TTI regime}
\label{TTI}

In the following we will make extensive use of this aforementioned timescale separation, focusing first on our analytical derivation in the time-translational invariant regime.
We consider the time evolution of the  spin variable $s(t)$ written in terms of an effective Langevin process:
\begin{equation}
\dot{s}(t)= -\frac{\partial V(s(t))}{\partial s}+ \frac{p(p-1)}{2} \int dt'' R(t,t'') C^{p-2}(t,t'')s(t'')+\xi(t) 
\label{dot-s2}
\end{equation}
where $V(s(t))$ stands for a generic potential, whereas $\xi(t)$ is a normally distributed coloured noise with zero mean and covariance defined by Eq. (\ref{noise-noise}).
We use timescale separation to decompose
the second term on the RHS of Eq. (\ref{dot-s2}), playing the role of a friction contribution: %Thanks to the decomposition of $C(t,t')\simeq C_\text{TTI}(t-t')+C_\text{A}(t,t')$ with $C_\text{TTI}(0)=q_d-q_1$, $C_\text{TTI}(\infty)=0$ and $C_\text{A}(t,t)=q_1 \ \ \forall t$,  %(or $C_\text{A}(t'/t)=1$ if $t'/t=1$), 
%it can be decomposed in two contributions:
 \begin{equation}
\int_0^t dt'' R(t,t'') C^{p-2}(t,t'')s(t'') \simeq
%\frac{p(p-1)}{2}
\int_{t-\tau}^{t} R_\text{TTI}(t-t') \left[ C_\text{TTI}(t-t')+q_1 \right]^{p-2}s(t')dt' + %\frac{p(p-1)}{2}
\int_{0}^{t-\tau} R_A(t,t') C_A^{p-2}(t,t') s(t') dt'
\end{equation}
where $\tau/t\sim o(1)$, the response function
$R_\text{TTI}(t-t') =\frac{1}{T}\frac{d}{dt'}C_\text{TTI}(t-t')$,
$R_\text{TTI}(t-t') \rightarrow 0$ as $t-t' \rightarrow \infty$, and $R_\text{A}(t,t')$ is uniformly small but when integrated over time leads to a finite contribution. 
By integrating the contribution that accounts only for the time-translation invariant regime by parts, we eventually obtain:
\begin{eqnarray}
\int_{t-\tau}^{t} R_\text{TTI}(t-t') \left[ C_\text{TTI}(t-t')+q_1 \right]^{p-2}s(t')dt' &\simeq& \frac{1}{T(p-1)}\left[(C_\text{TTI}(0)+q_1)^{p-1} s(t)- (C_\text{TTI}(\tau)+q_1)^{p-1} s(t-\tau)\right]+ \\
&-& \frac{1}{T(p-1)}\int_{t-\tau}^{t} \left[C_\text{TTI}(t-t')+q_1\right]^{p-1}
\dot{s}(t') %\frac{ d s(t')}{dt'}
dt'\ .
\end{eqnarray}
%where the last two pieces are boundary terms that can be rewritten according to an appropriate $1$RSB ansatz. For sake of simplicity, we consider the off-diagonal value $q_0=0$.
For very large $\tau$, even if still much smaller than $t$, $C_{TTI}(\tau)\simeq 0$ and the above equation becomes
%Gathering all contributions together, we can rewrite the first part as:
\begin{eqnarray}
\int_{t-\tau}^{t} R_\text{TTI}(t-t') \left[ C_\text{TTI}(t-t')+q_1 \right]^{p-2}s(t')dt'&\simeq&\frac{1}{ T(p-1)}
\left[q_d^{p-1}-q_1^{p-1}\right]s(t)+ \nonumber \\
&-&\frac{1}{ T(p-1)}
\int_{t-\tau}^{t} \Biggl \lbrace \left[ C_\text{TTI}(t-t')+q_1 \right]^{p-1} -q_1^{p-1} \Biggr \rbrace %\frac{ d s(t')}{dt'} 
\dot{s}(t') dt' 
\\ &=&
\frac{1}{ T(p-1)}
\left[q_d^{p-1}-q_1^{p-1}\right]s(t)- \frac{2}{ p(p-1)}
\int_{t-\tau}^{t} \nu_\text{TTI}(t-t')
%\frac{ d s(t')}{dt'}
\dot{s}(t') dt' \nonumber
%\\
%&=&
%\frac{1}{ T(p-1)}
%\left[q_d^{p-1}-q_1^{p-1}\right]s(t)+ \\
%&+&\frac{1}{ T(p-1)}
%\int_{t-\tau}^{t} \frac{ d }{dt'}\Biggl \lbrace \left[ C_\text{TTI}(t-t')+q_1 \right]^{p-1} -q_1^{p-1} \Biggr \rbrace s(t') dt'
\label{final_TTI}
\end{eqnarray}
%so that the original Eq. (\ref{dot-s2}) can be rewritten as
%\begin{equation}
%\frac{ d s(t) }{dt}=-\frac{\partial V(s)}{\partial s}+\frac{p}{2T}(q_d^{p-1}-q_1^{p-1})s(t)-\int_{0}^{t} \nu_\text{TTI}(t-t') \dot{s}(t') dt' +\frac{p(p-1)}{2} \int_{0}^t R_A(t,t')C_A^{p-2}(t,t')s(t') dt' + \xi(t)
%\end{equation}
where 
\begin{equation}
\nu_\text{TTI}(t-t')\equiv \frac{p}{2T}\left[ C_\text{TTI}(t-t')+q_1 \right]^{p-1} -\frac{p}{2T}q_1^{p-1} \ .
\end{equation} 
Recalling Eq. \eqref{noise-noiseA}, we therefore also get:
\begin{equation}
\langle \xi_\text{TTI}(t) \xi_\text{TTI}(t') \rangle
%=2T\delta(t-t')+\frac{p}{2}C(t,t')^{p-1}-\frac{p}{2}q_1^{p-1}
\simeq 2T\delta(t-t')+T\nu_\text{TTI}(t-t') \ .
\end{equation}
%and $%\langle \xi_\text{A}(t) \xi_\text{A}(t') \rangle_c=
%\langle \xi_\text{A}(t) \xi_\text{A}(t') \rangle\simeq\frac{p}{2}C_\text{A}(t,t')^{p-1}$ for $t-t' >t\gg 1$. \\
The original Eq. (\ref{dot-s2}) can thus be rewritten as %an effective stochastic process for the single variable $s(t)$
\begin{equation}
\dot{s}(t)\simeq-\frac{\partial V(s)}{\partial s}+\frac{p}{2T}(q_d^{p-1}-q_1^{p-1})s(t)-\int_{t-\tau}^{t} \nu_\text{TTI}(t-t') \dot{s}(t') dt' +\xi_\text{TTI}(t)+
h(t)
%\frac{p(p-1)}{2} \int_{0}^{t-\tau} R_A(t,t')C_A^{p-2}(t,t')s(t') dt' + \xi_A(t)
\label{motion}
\end{equation}
%where the first two terms on the l.h.s can be collected together and denoted as a total noise contribution in the fast regime, $\nu_\text{TOT}$. By contrast, 
where the terms accounting for the slow (aging) dynamical behavior have been embedded into what can be considered a slowly evolving
\emph{effective field} 
\begin{equation}
h(t)\equiv\frac{p(p-1)}{2} \int_{0}^{t-\tau} R_A(t,t')C_A^{p-2}(t,t')s(t') dt' + \xi_A(t) \ .
\label{field}
\end{equation}

Eq. \eqref{motion} shows that the original dynamical problem can be mapped into a stochastic process for the single variable $s(t)$ in the presence of friction and subject to a quasi stationary \emph{effective potential} 
\begin{equation}
\mathcal V (s,h(t))=V(s)-\frac{p}{4T}(q_d^{p-1}-q_1^{p-1})s^2-h(t)s \ .
\end{equation}
%Hence, within such a dynamical formalism, we are now able to write the resulting equation for an effective stochastic process, 
Such a new stochastic process will therefore be associated to a quasi-stationary conditional probability distribution, which is nothing but the Boltzmann-Gibbs distribution at a given external temperature $T$,
\begin{equation}
\begin{boxed}{
P(s \vert h(t))= \frac{1}{Z(h)} \exp \left[ -\frac{\mathcal V(s,h(t))}{T}\right] 
%\left[ -\frac{V(s)}{T} +\frac{p}{4T^2}(q_d^{p-1}-q_1^{p-1})s^2+ \frac{h (t)s}{T} \right] 
\label{P_sh}
}
\end{boxed}
\end{equation}
and to a quasi stationary free-energy obtained from the corresponding partition function $Z(h)$, $F(h(t))=-T\ln(Z(h))$.
%By integrating out the fast degrees of freedom, the spin  variable will satisfy a stationary distribution at fixed value of the field, which is nothing but the Boltzmann-Gibbs distribution whose \emph{effective Hamiltonian} is written in terms of a smooth random potential $V(s)$, plus the contribution coming from the effective quasi-static field $h$ coupled to the particle degree of freedom, and an additional boundary term that explicitly depends on the overlap difference $(q_d^{p-1}-q_1^{p-1})$ between the diagonal and the off-diagonal values for a $p$-spin interacting model.
Here, we have followed \cite{cugliandolo2000} where a similar procedure was used in 
to study the motion of a particle moving in a random potential and in contact with two thermal baths varying on very different timescales.
As also shown in \cite{cugliandolo2000} the solution of the full dynamics requires a detailed characterisation of the statistical properties of the quasi-static field $h(t)$. We discuss it for the present problem in the next section.

\subsection{Aging Regime}
\label{aging_1RSB}

In the aging regime we assume that correlation and response obey generalised FDT relations with violation parameter $x$ and effective aging temperature $T_\text{eff}=T/x$.
Moreover, we can replace the dynamical variable $s(t')$, in the integral of the first term of the slowly evolving field, with its average on the short times fluctuations, $\langle s(t')\rangle$, or equivalently its average over the distribution in Eq. (\eqref{P_sh}), which self-consistently depends on the field  and can be directly expressed in terms of the free-energy $F(h)$
\begin{equation}
    h(t)\simeq \frac{p}{2 T_\text{eff}} \int_{0}^{t} \frac{\partial}{\partial t'} \left( C_\text{A} ^{p-1} (t,t') \right) \langle 
    s(t') \rangle_{h(t')} 
    +\xi_A(t)=-\frac{p}{2 T_\text{eff}} \int_{0}^{t} \frac{\partial}{\partial t'} \left( C_\text{A} ^{p-1} (t,t') \right) \frac{\partial F(h)}{\partial h(t')} 
    +\xi_A(t) \ .
\end{equation}
Starting from this self-consistent equation on $h(t)$ it is possible to show \cite{cugliandolo2000} that the corresponding slow non-Markovian dynamics coupled to a bath of temperature $T_\text{eff}$ is associated to the following stationary distribution for $h$
%According to the Cugliandolo-Kurchan ansatz \cite{cugliandolo2000} and the results of the previous Section, $P(h)$ can be written in the following form
\begin{equation}
%\begin{split}
P(h) %=&\frac{1}{Z} \exp \left[  -\frac{h^2}{ 2 T_\text{eff} \;  \nu_A(t-t')} -\frac{1}{T_\text{eff}} F(h)\right]=\\
%&
= \frac{1}{Z} \exp \left[ -\frac{h^2}{ 2 \left( \frac{p}{2} q_1^{p-1} \right)} - \frac{x}{T} F(h) \right] \ ,
%\end{split}
\end{equation}
where $Z$ is the the normalization factor.

%To go from the first to the second line, we have considered the underlying relationship between the effective temperature in a dynamical approach and the breaking parameter $x$ of the Parisi ansatz, i.e. $T_\text{eff} \leftrightarrow 1/(\beta x)$.
%We skip all mathematical details here since we will extensively cover this argument in Sec. (\ref{connection_st_dyn}).
The stationary distribution of the slowly evolving field can now be used to explicitly characterise the full probability distribution of the degrees of freedom: 
\begin{equation}
    P(s)=\int dh P(s|h)P(h)
\end{equation}
and of its moments.
Except for the parameter $x$, which will be determined and discussed in the next two section, we can now obtain all quantities of interest for aging dynamics, in particular the overlaps $q_1$ and $q_d$:
\begin{equation}
\begin{boxed}
{
\overline{\langle s^2 \rangle }= \frac{1}{Z} \int dh \; e^{\left( -\frac{h^2}{ 2 \left( \frac{p}{2} q_1^{p-1} \right)} - \frac{x}{T} F(h) \right)} \left(\frac{\int_{-\infty}^{\infty} ds \; e^{-\frac{V(s)}{T} +\frac{p}{4T^2}(q_d^{p-1}-q_1^{p-1})s^2 +\frac{ hs }{T} } s^2 }{\int_{-\infty}^{\infty} ds \; e^{-\frac{V(s)}{T} +\frac{p}{4T^2}(q_d^{p-1}-q_1^{p-1})s^2 +\frac{ hs }{T} }} \right) \equiv q_d \ ,
}
\end{boxed}
\label{s2}
\end{equation}
\begin{equation}
\begin{boxed}
{
\overline{\langle s \rangle^2 }= \left[ \frac{1}{Z} \int dh \; e^{\left( -\frac{h^2}{ 2 \left( \frac{p}{2} q_1^{p-1} \right)} - \frac{x}{T} F(h) \right)}  \left( \frac{\int_{-\infty}^{\infty} ds \; e^{-\frac{V(s)}{T} +\frac{p}{4T^2}(q_d^{p-1}-q_1^{p-1})s^2 +\frac{ hs }{T} } s}{\int_{-\infty}^{\infty} ds \; e^{-\frac{V(s)}{T} +\frac{p}{4T^2}(q_d^{p-1}-q_1^{p-1})s^2 +\frac{ hs }{T} }} \right) \right]
^2 \equiv q_1 
}
\end{boxed}
\label{s1}
\end{equation}
Remarkably, these equations do not involve the dynamics any longer and resemble the static ones obtained by the replica method. This is not a coincidence, and it is at the basis of the correspondence between dynamic and static approaches, as discussed below in more details.

\subsection{Relationship with the statics}
\label{connection_st_dyn}

Replica aficionados will certainly realise that the two equations above coincides with the ones that one obtain by the replica method for the overlaps within pure states (whether the phase is 1RSB or FRSB). In the following we illustrate this relationship in the simple $p=2$ case and assuming a 1RSB Ansatz. The generalization to larger values of $p$ and to a FRSB Ansatz is straightforward. 

At equilibrium the usual way to obtain equations for the overlap parameters and the effective temperature is based on exploitation of the replica method \cite{MPV}, which allows one to compute the replicated free energy $f=-\lim_{N\rightarrow \infty}\frac{T}{N}\overline{\ln(Z)}$ by means of the following identity %in the $n \rightarrow 0$ limit:
\begin{equation}
\overline{\ln {Z}}= \lim_{n \rightarrow 0} \frac{ \overline{Z^n}-1}{n} \ .
\end{equation}
Instead of dealing with the disordered average of the logarithmic, %which appears to be quite painful, 
one has just to compute and average the replicated partition function with $n$ distinct copies of the original system. \\
Generically, the replicated partition function can be eventually expressed in terms of an action $\mathcal{S}$ which is a function of the overlap matrix $Q_{ab}$:
\begin{equation}
\overline{Z^n}= \int \prod_{(ab)} \frac{d Q_{ab}}{\sqrt{2 \pi}} e^{\mathcal{S}[Q_{ab}]}
\end{equation}
given the usual definition of the overlap between two spin configurations labeled by the replica indices $a$, $b$:
\begin{equation}
Q_{ab} = \frac{1}{N} \sum \limits_i s_i^a s_i^b \ .
\end{equation} 
%In the large $N$ limit, the action $\mathcal{S}$ can be evaluated by the saddle point approximation to get predictions on the overlap, hence on the different emerging phases. In the spherical case, one can directly integrate over continuous variables and obtain a factor $\text{Tr} \ln{Q}$, while in the case of Ising spins, more attention is needed.
Different approximations can be introduced to correctly parametrise the overlap matrix $Q_{ab}$, the simplest being the replica symmetric (RS) one.
\begin{figure}[h]
\center
\includegraphics[scale=0.52]{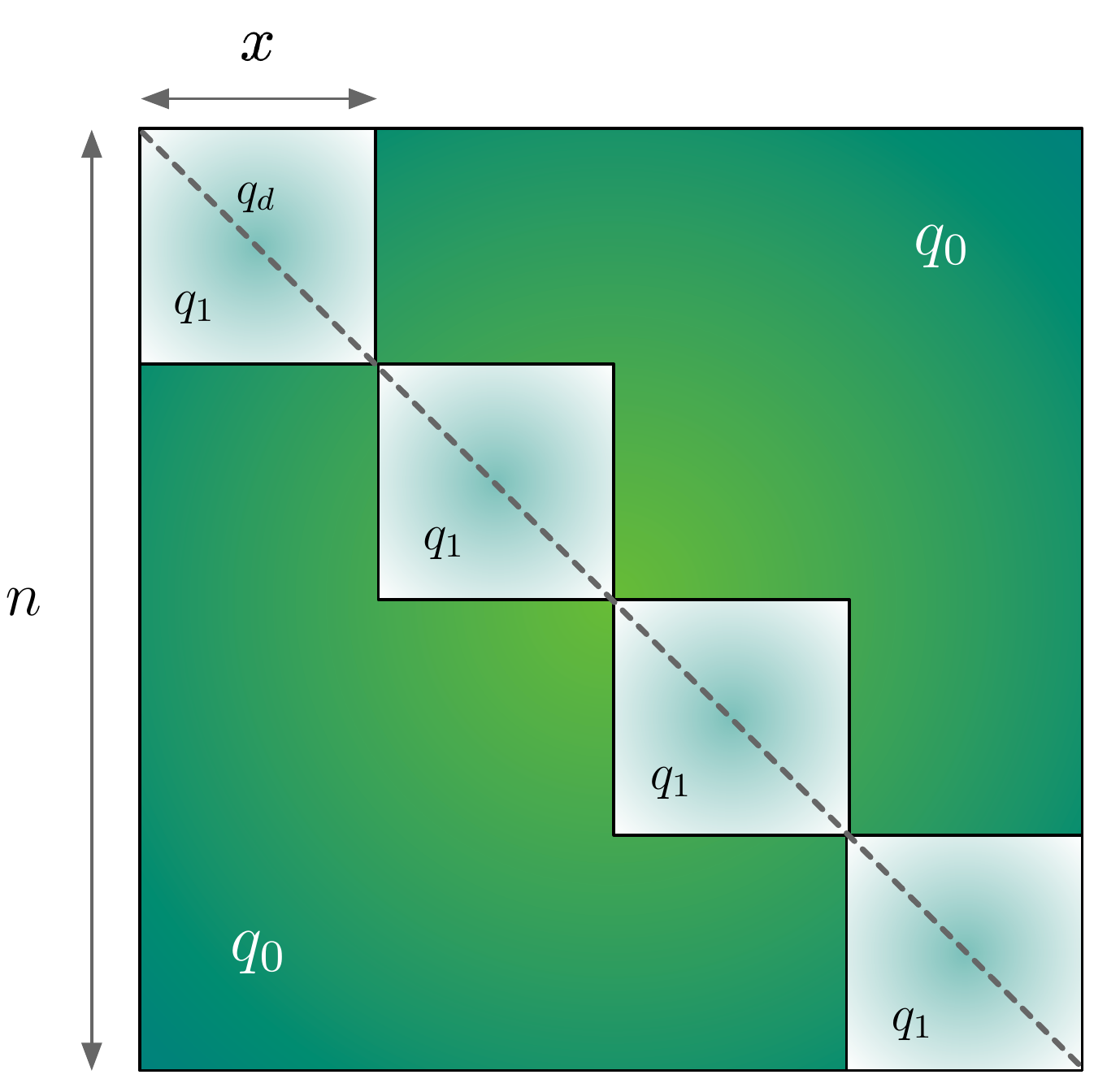}
\caption{Pictorial representation of a one-step replica symmetry breaking scheme for the overlap matrix $Q_{ab}$. The $n \times n$ matrix is divided into $n/x \times n/x$ blocks, each of them of size $x \times x$.
%The diagonal value, $q_d$, is fixed and equal to one both in the spherical $p$-spin and in the SK models, while the off-diagonal elements can be either equal to $q_1$ if the replica indices belong to the same block $\mathcal{B}$ (in light green) or to $q_0$ if the elements are outside the diagonal blocks (in dark green). This scheme can be iterated $k$ times and used to construct a $k$RSB scheme for the overlap matrix.
}
\label{block_matrix}
\end{figure}
This solution however is unable to describe the physics of disordered systems in the low-temperature regime, whose correct solution is based on an iterative block structure of that matrix \cite{Parisi1980, Parisi1983, Parisi1987, MPV}. Figure \ref{block_matrix} shows a representation of a $1$RSB realisation of such structure in an $n \times n$ matrix parametrised by
%However, once it was realized that the RS picture is unable to describe the physics of disordered systems in the low-temperature regime, several attempts have been tried to go beyond and take symmetry breaking effects into account. Now it is well known that the correct solution dates back to a series of works by Parisi based on an iterative scheme to be performed on that matrix \cite{Parisi1980, Parisi1983, Parisi1987, MPV}. 
a diagonal value $q_d$, and off-diagonal elements either equal to $q_1$ if the replica indices belong to the same block $\mathcal{B}$ of size $x\times x$ (in light green) or equal to $q_0$ if the elements are outside the diagonal blocks (in dark green). This scheme can be iterated $k$ times and used to construct a $k$RSB structure for the overlap matrix, which becomes Full RSB in the $k\rightarrow \infty$ limit. \\
Note that the correspondence in the notation between the parameters of the $1$RSB structure and the moments of the dynamical variable within the DMFT approach is not accidental and hints at the identification of the corresponding dynamic and static quantities. 
For instance, the static replica computation of the %To highlight the connection between our DMFT approach and the statics, we compute the 
overlap $q_1$ in the $1$RSB scheme %with simple pairwise interactions ($p=2$). 
is obtained by averaging single site variables from two replicas belonging
to the same block $\mathcal{B}$, with a weight given by the replicated action $\mathcal{S}$ %in the $1$RSB {\it ansatz} and 
rewritten introducing an auxiliary variable $z$ to decouple single replica integrals in the $1$RSB {\it ansatz}. 

For pairwise interactions ($p=2$), the final equation for $q_1$ reads:
 \begin{equation}
\begin{split}
q_1 = \langle s_a s_b \rangle 
%= & \frac{1}{Z} \int dh \; P(h) \left[ \int d s \; s \; e^{-\frac{1}{T} \left[ V(s)- h s -\frac{1}{2 T} (q_d-q_1)s^2 \right] }\right]=\\
= &\frac{1}{\overline{Z^n}}\int d z \; \frac{e^{-\frac{z^2}{2 q_1}}}{\sqrt{2 \pi q_1}} \left(\int d s \; s \; e^{-\frac{1}{T} \left[V(s)-  z s -\frac{1}{2 T} (q_d-q_1)s^2\right] }\right)^2 \left( \int d s  \; e^{-\frac{1}{T} \left[ V(s)- z s -\frac{1}{2T} (q_d-q_1)s^2 \right] }\right)^{x-2}  \times \\
& \times \left[ \int d z \frac{e^{-\frac{z^2}{2 q_1}}}{\sqrt{2 \pi q_1}}\left(\int d s  \; e^{-\frac{1}{T} \left[V(s)-  z s -\frac{1}{2 T} (q_d-q_1)s^2 \right]}\right)^{x}  \right]^{\frac{n}{x}-1} 
\end{split}
\end{equation}
with
\begin{equation}
    \overline{Z^n}=\left[ \int d z \frac{e^{-\frac{z^2}{2 q_1}}}{\sqrt{2 \pi q_1}}\left(\int d s  \; e^{-\frac{1}{T} \left[V(s)-  z s -\frac{1}{2 T} (q_d-q_1)s^2 \right]}\right)^{x}  \right]^{\frac{n}{x}} \ .
\end{equation}
Considering the analytical continuation $n \rightarrow 0$, the equation becomes:
\begin{equation}
q_1=\frac{\bigintss{ d z \;  \left(\frac{\left(\int d s \; s \; e^{-\frac{1}{T} \left[ V(s)- z s -\frac{1}{2 T} (q_d-q_1)s^2\right] }\right)}{\left(\int d s \;  e^{-\frac{1}{T}\left[ V(s)-  z s -\frac{1}{2 T} (q_d-q_1)s^2 \right]}\right)}\right)^2 \left( e^{-\frac{z^2}{2 q_1}-\frac{x}{T}  F(z)}  \right)}  }{\int d z \; e^{-\frac{z^2}{2 q_1}-\frac{x}{T}  F(z)}}
\label{q_1_final}
\end{equation}
with
\begin{equation}
    F(z)=-T\ln\left(\int d s  \; e^{-\frac{1}{T} \left[V(s)- z s -\frac{1}{2 T} (q_d-q_1)s^2 \right]}\right) \ ,
\end{equation}
which is formally equivalent to the equation for $q_1$ derived in the DMFT computation.
The same correspondence holds between the overlap on the diagonal of the $1$RSB matrix and the equal time correlation of the dynamical variable, also called $q_d$.  \\
%as the last term in parenthesis represents the normalization $Z(h)$. 
%The parameter $x$ , which is conjugated to the overlap $q$, comes out from the replica approach in a natural way by simply assuming a block-structured matrix: while $q$ represents the degree of similarity between different replica groups, $x$ denotes the actual size of each of these blocks.
Interestingly, comparing the DMFT and the replica equations, it also becomes evident a one to one correspondence between the dynamical aging field $h$ and the auxiliary variable $z$. An intuitive understanding of such correspondence
can be acquired by re-deriving the static equations through the cavity approach \cite{MPV, Mezard2003cavity}, 
where it clearly emerges that single spin variables  
%based on the thermodynamic study of 
%where it clearly emerges that single
%spin variables 
are effectively subject to random local fields $z$ characterised by the same non trivial distribution as the aging fields in the DMFT approach. \\ 
Finally %from the comparison between replica and DMFT equations 
this highlights the well known \cite{Cugliandolo1994off, Baldassarri1995, Cugliandolo-Kurchan-Peliti1997, Bouchaud1998} link between the FDT violation parameter and the $1$RSB parameter $x$, as they play a formally identical role in the static and dynamic equations for $q_d$ and $q_1$.

\subsection{DMFT for Ising and spherical $p$-spin models}
\label{applications}
In this Section, in order to provide simple examples 
and show connections with known results, 
we apply the formalism we developed 
to the spherical and the Ising cases introduced in Sec. \ref{SGmodels}.

\subsubsection{Ising $p$-spin model}
For an Ising $p$-spin model $V(s)$ is (formally) zero for $s=\pm1$ and infinite otherwise, hence $s^2=1$ and $q_d\equiv \overline{\langle s^2 \rangle}=1$.
Moreover we have 
\begin{equation}
P(s | h(t))= \frac{e^{\frac {h(t)   s}{T}}}{2 \cosh(h/T)}
\label{P_Ising}
\end{equation}
with $Z(h)=2 \cosh(h/T)$ 
and $F(h)=-T\ln (2\cosh(h/T))$
and for the only non trivial overlap 
\begin{equation}
q_1\equiv \overline{\langle s \rangle^2}= \overline{\left(\frac{e^{h/T}-e^{-h/T}}{2\cosh(h/T)}\right)^2}=\overline{\tanh^2(h/T)}=\frac{\int dh \; e^{%\left(
-\frac{h^2}{2\left(\frac{p}{2}q_1^{p-1}\right)}%-\frac{x}{T}F(h)\right)
}\tanh^2(h/T)[\cosh(h/T)]^x}{
\int dh \; e^{%\left(
-\frac{h^2}{2\left(\frac{p}{2}q_1^{p-1}\right)}%-\frac{x}{T}F(h)\right)
}[\cosh(h/T)]^x
}
 \ .
\label{moments_Ising}
\end{equation}    

\subsubsection{Spherical $p$-spin model}

To study the spherical $p$-spin model, we consider a soft spherical constraint implemented with the introduction of a quadratic potential $V(s)=\frac{\lambda s^2}{2}$ involving the spherical parameter $\lambda$. The conditional probability distribution for spin dynamical variables thus becomes
\begin{equation}
P(s \vert h(t))= %\sqrt{\frac{\left[\lambda-\frac{p}{2T}(q_d^{p-1}-q_1^{p-1})\right]}{2\pi T}} 
\frac{1}{Z(h)}
\exp \Biggl \lbrace { -\frac{\left[\lambda-\frac{p}{2T}(q_d^{p-1}-q_1^{p-1})\right]}{2T}\left(s-\frac{h}{\left[\lambda-\frac{p}{2T}(q_d^{p-1}-q_1^{p-1})\right]}\right)^2
+ \frac{h^2}{2T\left[\lambda-\frac{p}{2T}(q_d^{p-1}-q_1^{p-1})\right]}\Biggr \rbrace } \ ,
\end{equation}
with 
\begin{equation}
Z(h)= \sqrt{\frac{2\pi T }{\left[\lambda-\frac{p}{2T}(q_d^{p-1}-q_1^{p-1})\right]}} 
\exp \Biggl \lbrace {  \frac{h^2}{2T\left[\lambda-\frac{p}{2T}(q_d^{p-1}-q_1^{p-1})\right]}\Biggr \rbrace } \ ,
\end{equation}
from which the long and short time limit of the correlation function turn out to be respectively
\begin{equation}
%\begin{split}
q_1\equiv \overline{\langle s \rangle^2 }=  \overline{\left(\int ds P(s \vert h) s \right)^2}%=\Biggl \lbrace \frac{1}{\lambda-\frac{p}{2T}(1-q_1^{p-1})} \left[ \frac{p(p-1)}{2} \int_0^{t} R_A(t,t') C_A(t,t')^{p-2} s(t') dt'+ \xi_A(t) \right] \Biggr \rbrace^2
=  \overline{ \frac{h^2}{ \left[\lambda -\frac{p}{2T} (q_d^{p-1}-q_1^{p-1}) \right]^2} } 
%\end{split}
\label{squared_means}
\end{equation}
\begin{equation}
%\begin{split}
q_d\equiv \overline{\langle s^2 \rangle}=  \overline{\int P(s \vert h) s^2 }
%&= \Biggl \langle \frac{T}{ \lambda -\frac{p}{2T} (1-q_1^{p-1})} + \frac{1}{\left[ \lambda-\frac{p}{2T}(1-q_1^{p-1}) \right]^2} \left( \frac{p(p-1)}{2} \int_0^{t} R_A(t,t') C_A(t,t')^{p-2} s(t') dt'+ \xi_A(t) \right)^2 \Biggr \rangle_h =\\
=  \overline{\frac{T}{ \lambda -\frac{p}{2T} (q_d^{p-1}-q_1^{p-1})} +\frac{h^2}{\left[ \lambda-\frac{p}{2T}(q_d^{p-1}-q_1^{p-1}) \right]^2}  } \ .
%\end{split}
\label{squared_s}
\end{equation}
The equation on their difference $q_d-q_1$, evaluated at $q_d=1$, becomes a condition on the spherical parameter $\lambda$ to be imposed so that the spherical constraint is always satisfied during the dynamics
\begin{equation}
    \lambda -\frac{p}{2T}(1-q_1^{p-1})= \frac{T}{1-q_1} \ .
    \label{spherical-pspin}
\end{equation}
Finally, by using the above equation on the spherical parameter $\lambda$ and $q_d=1$ in Eq. \eqref{squared_means} it is possible to rewrite the equation on $q_1$
as follows
\begin{equation}
    q_1\left[\lambda-\frac{p}{2T}(1-q_1^{p-1})\right]^2=\left(\frac{1}{\frac{p}{2}q_1^{p-1}}-\frac{x}{T\left[\lambda-\frac{p}{2T}(1-q_1^{p-1})\right]}\right)^{-1}
\end{equation}
which after some passages becomes
\begin{equation}
    \lambda=T+\frac{p}{2T}\left(1-q_1^p(1-x)\right) \ .
    \label{q1-pspin}
\end{equation}
%We have derived thus far all relevant information for establishing a complete mapping between dynamics and statics. The only missing information at this level concerns the derivation of a general equation for determining the marginal stability.

As we will show in detail in Appendix \ref{virial_spherical}, the equation for the Lagrange multiplier can be found in an alternative way by introducing a virial equation, namely by multiplying every side of the equation of motion by $s$ and averaging over the associated stochastic process. 
 In the case of the Ising model, instead, the corresponding virial equation leads to an automatically satisfied condition that trivially corresponds to the normalization of the distribution $P(h)$. For a detailed explanation, we refer the interested reader to the Appendix.

\section{Effective temperature in the $1$RSB case}
\subsection{A diagrammatic approach}
\label{Dyson}

 As we have shown in the previous section, the dynamical aging Ansatz allows us to establish the equations satisfied by the dynamical overlaps. The effective temperature, however, remains unknown.  
We will now present a general approach that allows one to derive the equation for the effective temperature in models for which the dynamical $1$RSB Ansatz holds. Our procedure is based on the physical requirement that the dynamics in the TTI sector is marginal, \textit{i.e.} the relaxation to the plateau $q_1$ is power-law and not exponential. \\ 
Our starting point is  the stochastic Eq. (\ref{motion}), which  describes relaxation dynamics in the TTI regime.  \\
We shall use standard diagrammatic perturbation theory following the procedure developed for equilibrium critical spin-glass dynamics \cite{fischer1993spin}. 
Eq. (\ref{motion}) can be rewritten as %an effective stochastic process for the single variable $s(t)$
\begin{equation}
\int_{-\infty}^tR_0^{-1}(t-t')s(t')dt'=-\frac{\partial V(s)}{\partial s} +\xi_\text{TTI}(t)+ h(t) \ ,
%\frac{p(p-1)}{2} \int_{0}^{t-\tau} R_A(t,t')C_A^{p-2}(t,t')s(t') dt' + \xi_A(t)
%\label{motion}
\end{equation}
where $R_0^{-1}(t-t')$ has a simple expression in the Fourier domain ($\mathcal{F}_\omega$ denotes the Fourier transform):
\begin{equation}
R_0^{-1}(\omega)=-i\omega+\mathcal{F}_\omega \left( \partial_t \nu_{\text{TTI}} (t) \right) \ .
\end{equation}
We now present the method in the simplified case in which no field $h(t)$ is present, and then later we 
explain how to generalize it. The response function can be expressed in terms of the self-energy $\Sigma$ as
\begin{equation}
R_{\text{TTI}}(\omega)=\frac{1}{R_0^{-1}(\omega)-\Sigma(\omega)} \ .
\end{equation}
This Schwinger-Dyson equation is generically represented in diagrammatic theory using a straight line for $R_0$ and a dashed circle for $\Sigma$:
\begin{eqnarray}
\begin{split}
 \feyn{fcf} &= \feyn{ff + fpf + fpfpf + \cdots} \\
     &= \feyn{\frac{ff}{1-(fp)}}
     \end{split}
 \end{eqnarray}
The dashed circle corresponds to all self-energy diagrams generated when doing the perturbation theory in the couplings corresponding to $\frac{\partial V(s)}{\partial s}$. 

In cases where there are no conservation laws, as for the mean-field glassy systems we focus on, the response function decreases exponentially to zero at large times in the high-temperature ergodic regime. This behavior changes for marginal (or also critical \cite{fischer1993spin}) dynamics where instead one expects a power-law relaxation. Accordingly,  the behavior at small $\omega$ of $R^{-1}_{\text{TTI}}(\omega)$ is linear in the former case and power-law with an exponent less than one for marginal and critical dynamics. Hence, the condition encoding the existence of marginal dynamics is: 
\begin{equation}
    \lim_{\omega \rightarrow 0}\frac{\partial R^{-1}_{\text{TTI}}(\omega)}{\partial\omega}=\infty\,.
    \end{equation}
We now show that this requirement leads to a simple equation. In fact, using that at large times 
\begin{equation}
\partial_t\nu_{\text{TTI}} (t)\simeq \frac{p(p-1)}{2T}q_1^{p-2}\partial_t C_{\text{TTI}}(t)=-\frac{p(p-1)}{2}q_1^{p-2} R_{\text{TTI}}(t)\,\,,
\end{equation}
one finds that for marginal dynamics 
and in the small $\omega$ limit:
\begin{equation}
R_0^{-1}(\omega)\simeq-i\omega-\frac{p(p-1)}{2} q_1^{p-2} R_{\text{TTI}}(\omega) \ .
\end{equation}
By taking the inverse of the Schwinger-Dyson equation and differentiating it, one gets in the small $\omega$ limit:
\begin{equation}
\frac{\partial R^{-1}_\text{TTI}(\omega)}{\partial \omega}=\frac{\partial R_0^{-1}(\omega)}{\partial \omega}-\frac{\partial \Sigma(\omega)}{\partial \omega}\simeq\frac{\partial}{\partial \omega} \left[ -i \omega  -\frac{p(p-1)}{2} q_1^{p-2} R_\text{TTI}(\omega) \right] - \frac{\partial \Sigma(\omega)}{\partial \omega} \ .
\label{Gomega}
\end{equation}
Using the identity:
\begin{equation}
\frac{\partial R_\text{TTI}(\omega)}{\partial \omega} =-\frac{\partial R^{-1}_\text{TTI}(\omega)}{\partial \omega} R^2_\text{TTI}(\omega) 
\end{equation}
we finally obtain:
\begin{equation}
\frac{\partial R^{-1}_\text{TTI}(\omega)}{\partial \omega}=\frac{-i - \ \partial \Sigma(\omega) /\partial \omega}{1- \frac{p(p-1)}{2} q_1^{p-2} R^2_\text{TTI}(\omega)} \ .
\label{instability}
\end{equation}
It can be shown that the numerator is not singular (\textit{e.g.} to all orders in perturbation theory) \cite{fischer1993spin}. In consequence, the divergence for $\omega\rightarrow 0$ of the LHS --- the  condition for dynamical marginality --- is given by the vanishing of the denominator: 
\begin{equation}
1= \frac{p(p-1)}{2} q_1^{p-2} \left.R^2_\text{TTI}(\omega)\right|_{\omega=0}\ .
\end{equation}

When a random field $h$ is present in the stochastic equation, one has to redo the previous procedure introducing a $h$-dependent response function $\tilde{r}(\omega, h)$, which when averaged over the static field $h$ leads to the average response function: $\overline{ \tilde{r}(\omega, h)}= R(\omega)$. By repeating the previous analysis, see \cite{fischer1993spin} for the similar case of critical dynamics, one finds:
\begin{equation}
1= \frac{p(p-1)}{2} q_1^{p-2} \overline{ \tilde{r}^2(0,h) } 
\label{static_g}
\end{equation}
By using FDT, which is valid in the TTI regime, one obtains 
$\overline{ \tilde{r}^2(0,h)}= \overline{ \left(\frac{\langle s^2 \rangle -\langle s \rangle^2}{T} \right)^2 }=\overline{\left(\frac{\partial \langle s\rangle}{\partial h}\right)^2}$
and hence a condition for marginal dynamics that depends only of the probability distribution of $h$ and which therefore provides the extra equation allowing to fix 
the effective temperature: 
\begin{equation}
1= \frac{p(p-1)}{2} q_1^{p-2}  \overline{ \left(\frac{\langle s^2 \rangle -\langle s \rangle^2}{T} \right)^2 } \ .
\label{marginalcond}
\end{equation}
It can be shown that this is exactly the expression for the vanishing of the replicon in the $1$RSB analysis of this model. This therefore completes the analysis of the aging dynamics in the $1$RSB case, and shows how one can establish the connection with the static formalism. \\
As before, in order to show a simple application and the connection with known results, we apply the result above to i) Ising spins and $p \ge 2$, ii) continuous variables and $p>2$.
\vspace{0.3cm} 

\subsection{Applications to Ising and spherical $p$-spin models with $p>2$}
\subsubsection{Ising $p$-spin model}

Application of Eq. \eqref{marginalcond} to the Ising $p$-spin model requires the use of the previously derived Eq. %(\ref{P_Ising})-
(\ref{moments_Ising}) and $q_d=\langle s^2\rangle=1$ to get
\begin{equation}
\overline {\tilde{r}^2(0,h)}= \overline {\left(\frac{\langle s^2\rangle-\langle s\rangle^2}{T}\right)^2}= \frac{1}{T^2} \overline{  \left[ 1-\tanh^2( h/T) \right]^2 } \ ,
\end{equation}
and therefore 
\begin{equation}
1= \frac{p(p-1)}{2 T^2}  q_1^{p-2}  \left[1-2 \overline{ \tanh^2(h/T) } +\overline{ \tanh^4(h/T)  }\right]  \ ,
\label{Isingmarginal}
\end{equation}
which coincides with the expression derived in \cite{Crisanti2005} (see also \cite{Rizzo2013}). \\
The last condition, together with the Eq. \eqref{moments_Ising} on $q_1$ and $q_d=1$, forms a closed system of equations derived in this case within the DMFT approach, which can be therefore used to determine the $q_1$ and $x$ that characterise aging dynamics for an Ising $p$-spin with $p>2$. 
For $p=2$ the situation will be different as a Full RSB phase is going to control the aging behaviour of a relaxation dynamics after a quench. In this case a specific extension of DMFT must be considered as explained in Sec. \ref{full}. 
In this case, Eq. \eqref{Isingmarginal} evaluated at $p=2$, together with the condition $x=1$, can be used to determine at what temperature aging dynamics would set in.
%in particular of the long-time limit of the self-correlation function in the TTI regime:
%\begin{equation}
%q_1=\overline{\langle s \rangle_h^2 }= \overline{ \left[\tanh(h/T)\right] ^2 }\ .
%\end{equation}
%we can immediately compute, according to Eq. (\ref{static_g}), the static propagator $\tilde{r}(0,h)$. It is simply obtained by the differentiation of the average spin value with respect to the effective field, namely:
%\begin{equation}
%\langle \tilde{r}^2(0,h)\rangle_h= \overline{\left( \frac{\partial \langle s \rangle }{ \partial h} \right)^2}=  \overline{\left( \frac{\partial \tanh(h/T)}{\partial h} \right)^2 } = \frac{1}{T^2} \overline{  \left[ 1-\tanh^2( h/T) \right]^2 } \ ,
%\end{equation}
%which implies that the following condition has to be satisfied if the system is marginally stable:
%\begin{equation}
%\begin{boxed}{
%1= \frac{p}{2 T^2} (p-1) q_1^{p-2} \left[1-2 \overline{ \tanh^2(h/T) } +\overline{ \tanh^4(h/T)  }\right]}
%\end{boxed}
%\label{replicon_p_Ising}
%\end{equation}
%which precisely coincides with the expression derived in \cite{Crisanti2005}.
%\vspace{0.7cm}
%\item {\textbf{Spherical $p$-spin}}\\
\subsubsection{Spherical $p$-spin model}
To obtain a similar closed set of equations in the spherical case, we recall again the results of Sec. (\ref{applications}) and in particular Eqs. \eqref{q1-pspin} and
\eqref{spherical-pspin}.
The addition of Eq. \eqref{marginalcond}, which in this case gives 
\begin{equation}
1= \frac{p(p-1)}{2 } q_1^{p-2}  \overline{ \left(\frac{\langle s^2 \rangle -\langle s \rangle^2}{T} \right)^2 }=\frac{p(p-1)}{2 T^2} q_1^{p-2}  (1-q_1)^2 \ ,
\label{replicon_spherical}
\end{equation}
closes the set so that it is possible to determine $\lambda$, $q_1$ and $x$ within the DMFT approach.
Note that this set of equations corresponds to the one obtained in the works on the spherical $p$-spin model \cite{Crisanti1992statics, Cugliandolo1993}.

\section{Aging dynamics in the Full RSB regime}
\label{full}
The aim of this section is to show how to tackle cases with an infinite number of slow time-scales. For the class of models we focus on, this happens for Ising spins and $p=2$, which corresponds to the Sherrington-Kirkpatrick (SK) model.
The different nature of its transition (spin-glass like) largely affects the kind of aging behaviour taking place in the long time dynamics. We have therefore to derive a new rule for the slow evolution of the external field.
Conversely the description of the short time dynamics within a TTI framework will remain unchanged. 
In particular the dynamical variable $s(t)$ on short time scales evolves according to a stochastic process in the presence of friction as described by Eq. \eqref{motion} with $q_d=1$, 
\begin{equation}
    \nu_{\text{TTI}}(t-t')=\frac{1}{T}C_\text{TTI}(t-t') \ ,
\end{equation}
(since $p=2$), 
and associated to a quasi stationary conditional probability $P(s|h(t))$ of the form in Eq. (\ref{P_sh}), 
which in this case becomes \begin{equation}
P(s \vert h(t))= \frac{e^{\frac{h(t)   s}{T}}}{2 \cosh(h(t)/T)} \ ,
\end{equation}
and immediately implies $\langle s(t)\rangle=m(t)= \tanh( h(t)/T)$.
Recall that the slowly evolving external field $h(t)$ was defined in Eq. (\ref{field}) and in this case given by 
\begin{equation}
    h(t) = \xi_\text{A}(t)+\int_0^tdt'R_\text{A}(t,t')m(t')
\end{equation}
where $\langle\xi_\text{A}(t)\xi_\text{A}(t')\rangle=C_\text{A}(t,t')$.
Finally, the slow evolution of such external field, controlled by aging dynamics, will set in as soon as the condition of marginal stability in Eq. (\ref{Isingmarginal}) for $p=2$ is satisfied:
\begin{equation}
0=1-\frac{1}{T^2}  \left[1-2 \overline{ \tanh^2(h/T) } +\overline{ \tanh^4(h/T)  }\right] \ .%=1-\frac{1}{T^2} \int dh\;  P(h )\left[ 1- \tanh(h/T)^2 \right]^2 \ .
\label{replicon_SK}
\end{equation}
In the following we derive an explicit equation that describes the evolution of the aging field $h(t)$ along the dynamics.
To this aim we notice that 
\begin{equation}
\begin{split}
 %h(t+\Delta t)= &\xi_\text{A}(t+\Delta t) + \int_{0}^{t + \Delta t} dt' \; R_\text{A}(t+\Delta t,t') m(t') \\
 \Delta h(t)=&h(t+\Delta t)-h(t)=  \xi_\text{A}(t+\Delta t)-\xi_\text{A}(t) + \int_{0}^{t+\Delta t} dt' R_\text{A}(t+\Delta t,t') m(t')- \int_{0}^{t} dt' R_\text{A}(t,t') m(t')\\
 =&\xi_\text{A}(t+\Delta t)-\xi_\text{A}(t)+ \left[ R_\text{A}(t,t) m(t) +\int_{0}^{t} \partial_t R_\text{A}(t,t') m(t') d t' \right] \Delta t \ .
 \label{deltah}
\end{split}
\end{equation}
Note that $\Delta t$ represents a small change in unit of very large time-scales. 
There are three contributions to the change of the slow field: the first is due to the evolution of the stochastic slow noise between $t$ and $t+\Delta t$, 
the second depends on the state of the system at time $t$, and the last is obtained integrating over all the past behavior. 

By dropping the last term, which gives a sub-leading contribution, one can recognize that the equation on $\Delta h(t)$ has the form of a stochastic equation with a drift term: 
\begin{equation}
    D^{(1)}(t)=  R_\text{A}(t,t)m(t)\Delta t ={{-\left.\frac{x}{T}\frac{\partial C_\text{A}(t,t')}{\partial t}\right|_{t'=t} m(t) \Delta t 
    \simeq \frac{x}{T}[ C_\text{A}(t,t)-C_\text{A}(t+\Delta t,t)] m(t) }} 
    %\left[ \frac{X(t,t)}{T} \frac{\partial C_\text{A}(t,t)}{\partial t}m(t) \right]}
\end{equation}

\begin{figure}
\includegraphics[scale=0.45]{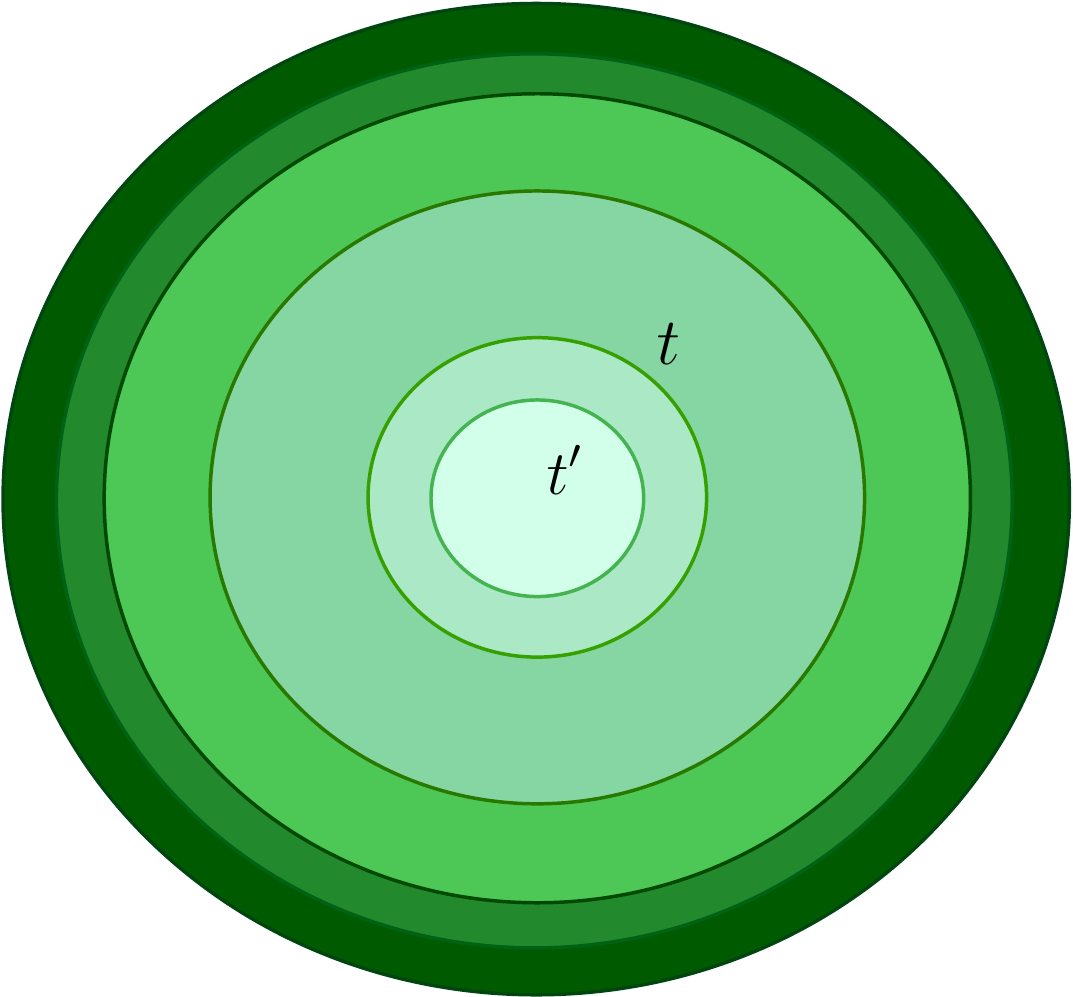}
\caption{Representation of $k$ different time sectors (for simplicity with $k=6$) showing the hierarchical organization of timescales according to a $k$-RSB Ansatz. The Full RSB picture corresponds to sending the number of sectors $k$ to infinity. In this limit, the piecewise function parametrising the overlap matrix is well defined and converges to the continuous function $q(x)$. }
\label{time_sec}
\end{figure}

and a noise term with variance that reads (keeping only terms up to the linear order in $\Delta t$):
\begin{equation}
\begin{split}
    D^{(2)}(t)=& 
    {{\overline{\xi_\text{A}(t+\Delta t)\xi_\text{A}(t+\Delta t)}+\overline{\xi_\text{A}(t)\xi_\text{A}(t)}-\overline{\xi_\text{A}(t+\Delta t)\xi_\text{A}(t)}-\overline{\xi_\text{A}(t)\xi_\text{A}(t+\Delta t)} }}
    %=(\partial_tC_\text{A}(t,t')-\partial_{t'}C_\text{A}(t,t'))\Delta t 
    \\
    =& C_\text{A}(t+\Delta t,t+\Delta t) +C_\text{A}(t,t)-2C_\text{A}(t+\Delta t,t)
    \ .%\left[ \frac{X(t,t)}{T} \frac{\partial C_\text{A}(t,t)}{\partial t}m(t) \right]
\end{split}
\end{equation}
In order to evaluate the differences of correlations in the drift and the variance of the noise we use the FRSB aging Ansatz discussed in Section \ref{1timescale}.  
The aging correlation function $C_A(t,t')$ with $t\geq t'$ equals the intrastate overlap $q(x)$ of the states reached at the largest timescale $t$ and associated to FDT violation parameter $x$. 
%The variable $x\in [0,1]$ identifies the states according to their position in the Full RSB hierarchy and in this sense we expect that $q(x)$ eventually corresponds to the Parisi Full RSB order parameter.
%The larger is the time scale of the dynamics, the smaller the corresponding interstate correlation (or overlap), and the more distant is the FDT violation parameter from $1$, {\it i.e.} from no violation. By this scheme if, for $t\geq t'$, we equate $C_\text{A}(t,t')=q(x)$, we also have $C_\text{A}(t+\Delta t,t')=q(x-\Delta x)$ and $C_\text{A}(t,t')-C_\text{A}(t+\Delta t,t')=q(x)-q(x-\Delta x)\simeq \dot{q}(x)\Delta x$.
Therefore the drift and the variance can be rewritten in terms of the intrastate overlaps as follows
\begin{equation}
    D^{(1)}(t)\simeq \frac{x}{T} m(h) \left(q(x)-q(x-\Delta x)\right)
    \simeq\frac{x}{T} m(h)\dot{q}(x)\Delta x \
\end{equation}
\begin{equation}
    D^{(2)}(t)\simeq q(x-\Delta x)+q(x)-2q(x-\Delta x)
    %=q(x)-q(x-\Delta x)
    \simeq \dot{q}(x)\Delta x \ .
\end{equation}
Finally we note that the covariance of the noise at different timescales is zero since: 
\begin{equation}
   \overline{\xi(t+\Delta t)\xi(t'+\Delta t)}+\overline{\xi(t)\xi(t')}-\overline{\xi(t+\Delta t)\xi(t')}-\overline{\xi(t)\xi(t'+\Delta t)}= q(x-\Delta x)+q(x)-q(x-\Delta x)-q(x)=0\ .
\end{equation}

In conclusion we have obtained that the slow field $h(t)$ satisfies a stochastic Langevin equation
\begin{equation}
\begin{boxed}{
\frac{d h(x)}{dx}= \frac{x}{T}  m(h)\dot{q}(x)   + \sqrt{\dot{q}(x)} z(x) \ .}
\end{boxed}
\label{stoc_frsb}
\end{equation}
where $z(x)$ is a Gaussian white noise,
$\overline{z(x)z(x')}=\delta(x-x')$, and the evolution is measured in terms of the change of the effective temperature $x$ (each $x$ corresponds to a time-scale $t_x$ as recalled in Fig. \ref{time_sec}).  
\\

Remarkably, such Langevin's equation coincides with the one derived in the studies that focused on 
the thermodynamic FRSB phase \cite{MPV, Mezard1985microstructure, Crisanti-Rizzo2002, Parisi2017marginally,Sommers1983, Sommers-Dupont}. This shows that the distributions of the effective fields on the slow time-scales coincide with the ones of the thermodynamic FRSB solution in the hierarchical clusters. Moreover,
also $m(h)$ and $q(x)$ have the same expression than in the static case: 
$m(h)$ is the magnetization on the time-scale $t_x$, hence it is averaged over all effective fields corresponding to smallest time-scales and it depends on $h(x)$, which is the effective field at time $t_x $. Whereas $q(x)$ is the overlap on the time-scale $t_x$, {\it i.e} it is the square of the magnetization $m(h(x))$ average over $h(x)$.
In order to solve equation \ref{stoc_frsb}, one has to know $m(h)$ and $q(x)$ which are determined self-consistently from the solution of the stochastic equation itself.  
Given the one-to-one mapping with the static case, we refer to the classic original paper on FRSB for more details \cite{MPV, Mezard1985microstructure,Sommers1983, Sommers-Dupont}.

In summary, our procedure shows (as it was expected from the solution of simplifed models \cite{Cugliandolo1994,Franz1994}) that aging dynamics in the FRSB case is strongly related to the static solution. Indeed, we have provided a purely dynamical derivation 
of the stochastic equation (\ref{stoc_frsb}) at the basis of FRSB.

\subsection{Relationship and analogies with Sompolinsky's dynamical approach}
\label{stat&sompolinsky}

The first who proposed a deep investigation of the spin glass phase using a dynamic approach was Sompolinsky in the eighties \cite{sompolinsky1981time, Sompolinsky1982}. He proposed a way to obtain the properties of the spin-glass phase using an approach that takes into account dynamics over diverging time-scales. Cutting a long story short, his main assumptions are:
\begin{itemize}
    \item  there exists an infinite number of diverging timescales belowe $T_c$ in mean-field spin-glasses.
    \item the spin-spin correlation function is affected by all those timescales, in particular for each time-scale $t_x$, one gets  
\begin{equation}\label{som1}
q(x)=\overline{\langle s_i(0) s_i(t_x) \rangle} \ .
\end{equation}
\item the equilibrium response function until the diverging timescale $t_x$ 
is given by:
\begin{equation}
    \chi(x)=\int_0^{t_x}R(t_x,t')dt'=\frac{1}{T}\left(\Delta(x)+(1-q(1))\right)
\end{equation}
where the first term on the right hand side is by definition the contribution to the response from the diverging time-scales, and the second term is the contribution due to the fast degrees of freedom ($q(1)$
is related to the spin-spin correlation, see Eq. (\ref{som1})).

\item The anomalous part of the response and the correlation on diverging timescales are related by 
 \begin{equation}
\dot{\Delta}(x)= - x \dot{q}(x) \ .
\label{Gauge}
\end{equation}
\end{itemize}
The physics behind Sompolinsky's solution was never fully justified; nevertheless it was shown that these assumptions 
allows to recover the Parisi solution of spin-glasses. 
\\

In the following we show that the last assumption---the crucial one---is analogous to the generalized fluctuation-dissipation relation we used in the last section. In fact, within the dynamical aging ansatz the slow time-scales lead to an anomalous contribution to the aging response function for $t,t' \gg 1$:
\begin{equation}
   \int_{t'}^{t}R(t,t'')dt''=\frac{1-q(x_{max})}{T}+\int^{x_{max}}_{x}\frac{x}{T}\dot{q}(x')dx'
\end{equation}
where we have used the same notation of the previous section, and $t,t'$ are taken in the time sector corresponding to $x$, i.e. $0<\frac{h_x(t')}{h_x(t)}<1$ ($x_{max}$ is the value of $x$ corresponding to the first plateau in the correlation function, and we have traded the index $i$ in $h_i(t)$ for the corresponding value of $x$).\\
Identifying the response in the LHS above with $\chi(x)$, and taking the derivative with respect to $x$ we discover that the Sompolinsky's relation  $\dot{\Delta}(x)= - x \dot{q}(x)$ is mutatis mutandis the generalized fluctuation-dissipation relation. \\

The interpretations of our and Sompolinsky approaches are clearly different: we study 
aging dynamics whereas he wasn't discussing off-equilibrium. However, algebraically the two approaches are identical, and the assumption $\dot{\Delta}(x)= - x \dot{q}(x)$ becomes 
under the lens of the off-equilibrium approach the generalized FDT discovered by Cugliandolo and Kurchan in their study of mean-field aging \cite{Cugliandolo1993}. This result offers a new perspective on the Sompolinsky's solution, and clarifies its algebraic equivalence with the Parisi's solution of spin-glasses.

\section{Conclusions and Perspectives} 
\label{conclusions}

Developing a mean-field procedure to study the dynamics of out-of-equilibrium systems has been a fundamental step in the theory of disordered and amorphous systems. It has allowed to address challenging questions about low-temperature glassy behaviors, and to understand the role of complex landscapes in determining slow dynamics \cite{Bouchaud1998, Cugliandolo2002LH, Cugliandolo2003slow, Castellani2005}. 

In this work, we have considered cases in which the slow dynamics  
is studied by Dynamical Mean-Field Theory (DMFT), and the resulting equations do not always simplify in integro-differential equations on response and correlation functions. 
Instead, one has to deal with the full-fledged self-consistent problem in which the thermal bath properties are determined from the stochastic process induced by the bath itself.  
Our approach is based on the mean-field theory of aging dynamics \cite{Cugliandolo1994_aging, Bouchaud1998, Franz2000, Cugliandolo2002LH, Cugliandolo2003slow, Biroli2005}.
It relies on the hypothesis of well-defined timescale separation, between fast degrees of freedom, leading to a time translational invariant (TTI) regime, and slow degrees of freedom, leading to an aging regime. 
This separation of time scales feeds into the self-consistent stochastic process 
associated to DMFT: it leads to generalized friction and noise that also have  fast and slow contributions. Our main result is establishing a procedure
that allows to study this self-consistent dynamical problem and obtain the main quantities of interest, e.g. effective temperatures, correlations and responses on slow time-scales. The resulting equations make explicitly the link between aging dynamics and static replica computations, which was worked out in simplified models \cite{Cugliandolo1993, Cugliandolo1994off, Sommers1983, MPV, Franz1994, Barrat1997, Rizzo2013} and then assumed to valid more generally. They also display strong relationship with the quasi-equilibrium picture of glassy dynamics \cite{Franz2013, Franz2015quasi}.

A natural extension of our work concerns the so-called Gardner phase \cite{Gardner1985}, which appears in many mean-field models at low temperature, and it has been shown to have very important consequences in jamming physics \cite{Charbonneau2014}. 
 The associated dynamical behavior is expected to be a combination of the $1$RSB and Full RSB studied in this paper, so our results provide a good starting point to address it. Recent works have unveiled that the \emph{weak long-term memory} property at the basis of the mean-field theory of aging is not verified at least in certain models and for certain initial conditions (quenches from finite temperature) \cite{Folena2019,Bernaschi2019}. It would be interesting to investigate how our framework can be generalised to these cases.

Let us conclude highlighting possible direct applications of our findings. We foresee three main directions:
\begin{itemize}
    \item {\it Dynamical theory of aging and shear of glasses in the limit of infinite dimensions}. The DMFT equations to treat those cases have been established recently in \cite{Agoritsas2019,Agoritsas2019out}. Our framework provides a way to analyze them and generalise the results obtained 
on these topics using simplified mean-field disordered systems \cite{Cugliandolo1993, Berthier2002}. 
\item {\it Dynamical theory of slow dynamics in well mixed interacting ecosystems}. Simple models of ecosystems with a large number of species \cite{Bunin2017, Barbier2018, RoyDMFT} have been shown to display aging and, in cases of symmetric interactions, Full RSB physics \cite{Biroli2018, Altieri_eco2020}. Since DMFT naturally applies to them \cite{RoyDMFT, Galla_dynamics}, a generalization of our approach offers a promising way to develop a theory of such phenomena.  
\item {\it Inference and Machine Learning}. Gradient descent algorithms are natural methods to deal with non-convex optimization problems. Although they are widely used in practice, a theory of their algorithmic threshold is lacking. Only very recently a first result has been obtained in a model of matrix-tensor PCA \cite{Sarao,Sarao2020}. The study of the gradient descent dynamics in models such as the non-convex spherical perceptron \cite{Franz2016}, and generalized linear models \cite{maillard2019} can be done by DMFT \cite{Agoritsas2018}. In consequence, our approach combined with the one of \cite{Sarao2020, Sarao} provides a guide to develop a theory of the algorithmic threshold of gradient descent in these contexts. 

\end{itemize}

\section*{Acknowledgments}
We thanks L. Cugliandolo and J. Kurchan for helpful discussions. 
This work was supported by the Simons Foundation Grants No. $\#454935$ (Giulio Biroli).

\vspace{0.5cm}

\appendix

\section{Alternative analysis of the spherical p-spin in terms of the virial equation} 
\label{virial_spherical}

\subsection{1st approach}
We propose here a complementary approach to derive the expression of the Lagrange multiplier in the spherical $p$-spin model.
Our starting point is the equation for the spin evolution, Eq. (\ref{dot-s2}) of the main text, in which we multiply both sides by $s$ and average over the stochastic process. In this way, we can express the expectation value of the constraining force term in an analytically treatable way.
\begin{equation}
\biggl \langle s \frac{ds}{dt} \biggr \rangle=-\biggl \langle s \frac{ \partial V}{\partial s} \biggr \rangle + \frac{p(p-1)}{2} \int_{0}^{t} R(t,t') C^{p-2}(t,t')\langle s(t)s(t') \rangle dt'  +\langle \xi(t) s(t) \rangle 
\label{eq_motion1}
\end{equation}
Using a property of Gaussian integrals, we can simplify the term $\langle \xi(t) s(t) \rangle$ by means of \emph{Novikov theorem} \cite{Camacho2013} (aka \emph{Girsanov theorem} in mathematical jargon).
Keeping the discussion as general as possible, let us suppose to be interested in computing the following value
\begin{equation}
\begin{split}
\langle \phi_k F(\phi) \rangle= & \frac{1}{Z_0[0]} \int d \phi \; \phi_k F(\phi) e^{-\frac{1}{2} \phi A \phi}= - \frac{1}{Z_0[0]} \sum_{n} \Delta_{k n} \int d \phi \; F(\phi) \frac{\partial}{\partial \phi_n} e^{-\frac{1}{2} \phi A \phi}=\\
=& \frac{1}{Z_0[0]} \sum_{n} \Delta_{kn} \int d \phi \; \frac{\partial}{\partial \phi_n} F(\phi) e^{-\frac{1}{2} \phi A \phi}
\end{split}
\end{equation}
where $\Delta=A^{-1}$ that in field theory corresponds also to the bare propagator, while the normalization factor is $Z_{0}[0]= \sqrt{2 \pi / \det{A}}$.
Using the aforementioned theorem, which remains valid as long as we consider Gaussian noises, the expectation value between the noise and the spin variable over the stochastic process can be rewritten in a more compact way as 
\begin{equation}
\langle \xi(t) s(t') \rangle =  \int dt'' \mathcal{M}(t,t'') R(t',t'') 
\end{equation}
where the kernel $\mathcal{M}(t,t')= 2 T\delta(t-t')+\frac{p}{2} C^{p-1}(t,t')$ contains a non-interacting part satisfying the TTI hypothesis, and an interacting non-translational invariant contribution.
%$ 2 T\delta(t-t')+\frac{p}{2} C^{p-1}(t,t')$.
Moreover, the first term on the LHS of Eq. (\ref{eq_motion1}), according to Ito's prescription, can be rewritten as
\begin{equation}
\frac{d s^2}{dt}= 2 s \frac{ ds}{dt} + 2T 
\end{equation}
which for a spherical model reduces to: $s \frac{ds}{dt}+T=0$.

Then, Eq. (\ref{eq_motion1}) becomes:
\begin{equation}
\begin{split}
& \biggl \langle s \frac{ds}{dt} \biggr \rangle=-\biggl \langle s \frac{ \partial V}{\partial s} \biggr \rangle + \frac{p(p-1)}{2} \int_{0}^{t} R(t,t') C^{p-2}(t,t')\langle s(t)s(t') \rangle dt'  +\langle \xi(t) s(t) \rangle \Rightarrow \\
& - T= -\lambda +\frac{p(p-1)}{2} \int_{0}^{t} R(t,t') C^{p-1}(t,t') dt'+ \int_{0}^{t} \left[ 2T \delta(t-t'')+\frac{p}{2} C^{p-1}(t,t'') \right] R(t,t'') dt''
\end{split}
\end{equation}
and, since $R(t,t)$ is zero for causality, it implies
\begin{equation}
-T=-\lambda +\frac{p^2}{2} \int_{0}^{t} R(t,t')C^{p-1}(t,t') dt' \ .
\end{equation}
This expression provides the well-known condition for the Lagrange multiplier in the case of a spherical $p$-spin model:
\begin{equation}
\begin{boxed}{
\lambda= T+ \frac{p^2}{2} \int_0^{t} R(t,t') C^{p-1}(t,t') dt'} \ .
\label{lagrange}
\end{boxed}
\end{equation}
If we integrate by part the argument in the integral we recover a compact expression for the spherical parameter:
\begin{equation}
\lambda=T+\frac{p}{2} \left( \frac{1-q_1^{p}}{T}+\frac{q_1^{p}}{T_\text{eff}} \right)
\end{equation}
which is equivalent to Eq. \eqref{q1-pspin}
and can be related to the expression of the asymptotic energy $\mathcal{E}_{\infty}$ as well, as known from Cugliandolo-Kurchan equations and corresponding to \cite{Cugliandolo1993}: 
\begin{equation}
\mathcal{E}_{\infty}=-\frac{1}{2 T} \left[ (1-q_1^p) + p q_1^{p-1} \int_{0}^{1} d \mu{''} \mathcal{R}(\mu{''}) \mathcal{C}^{p-1}(\mu{''}) \right]
\end{equation}
Note that the correlation and response depend now on the rescaled parameter $\mu \equiv t'/t$ that implies $C(t,t')=q \;\mathcal{C}(\mu)$ and $t R(t,t')= \mathcal{R}(\mu)$.

\vspace{0.5cm}

\subsection{2nd approach}

We again consider the equation of motion and multiply both sides by $s$
\begin{equation}
\biggl \langle s \frac{ds}{dt} \biggr \rangle= -\biggl \langle s \frac{ \partial V}{\partial s} \biggr \rangle + \frac{p(p-1)}{2} \int_{0}^{t} dt' \; R(t,t') C^{p-2}(t,t')\langle s(t)s(t') \rangle +
\langle \xi(t) s(t) \rangle 
\end{equation}
averaging then over the stochastic process.
At this level, we do not need to specify the precise form of the potential. We proceed integrating by parts the first term in the RHS to keep the computation as general as possible. The last expectation value can be treated exactly as before using \emph{Novikov's theorem}.
This term basically sums up to the friction term on the RHS leading to
\begin{equation}
-T=-\frac{1}{Z} \Biggl \lbrace \int ds \; \left[ -Ts \left ( \frac{ \partial}{\partial s} e^{-V(s)/T} \right) e^{\frac{p}{4T^2}(1-q^{p-1})s^2 +\frac{hs}{T}} \right] \Biggr \rbrace + \frac{p^2}{2} \int_0^{t} R(t,t') C^{p-1}(t,t') dt'
\end{equation}
\begin{equation}
\begin{split}
& -T=-\left[ T+\frac{p}{2T}(1-q_1^{p-1}) \langle s^2 \rangle_h + h \langle s \rangle_h \right] +\frac{p}{2 T} \left[1 -q_1^{p}(1-x) \right] \\
& \Rightarrow 0= -\frac{p}{2 T} (1-q_1^{p-1}) \langle s^2 \rangle_h - h \langle s \rangle_h +\frac{p}{2T} \left[ 1 -q_1^{p}(1-x) \right] \ .
\label{motion_virial_app}
\end{split} 
\end{equation}

In the spherical model, the field distribution $P(h)$ can be exactly computed and becomes
\begin{equation}
P(h) = \frac{1}{Z} \exp \left[ -\frac{h^2}{ 2\left( \frac{p}{2} q_1^{p-1} \right)} -\beta x F(h) \right]= \frac{1}{Z} \exp \Biggl \lbrace -\frac{h^2}{ 2\left( \frac{p}{2} q_1^{p-1} \right)} +\frac{ \beta x h^2}{2\left[ \lambda-\frac{p}{2T}(1-q_1^{p-1}) \right]} \Biggr \rbrace \ ,
\end{equation}
where  $T_\text{eff} = 1/(\beta x)$, while the normalization factor $Z$ is
\begin{equation}
Z=\frac{ \sqrt{2 \pi}}{\sqrt{ \frac{2 q_1^{1-p}}{p} +\frac{(-1+q_1) x \beta}{T} } } \ .
\end{equation} 
Using the additional condition on the spherical constraint, we can further simplify the denominator of the normalization factor and obtain a more compact expression for $\lambda$, as also reported in the main text:
\begin{equation}
\lambda-\frac{p}{2 T} (1-q_1^{p-1})= \frac{T}{1-q_1} \ .
\label{Lagrange_mult_app}
\end{equation}
Hence, the short and long-time limit of the correlation function, which correspond respectively to $q_d$ and $q_1$ in a static $1$RSB computation, are:
\begin{equation}
\overline{\langle s^2 \rangle}=   \frac{1}{Z} \int dh \; P(h) \left[ \frac{T}{T/(1-q_1)}+\frac{h^2}{\left( T/(1-q_1) \right)^2 } \right] =  \frac{(-1+q_1)\left[-2q_1 T^2-pq_1 ^p (-1+q_1)(-1+x)\right]} {T \left[ 2q_1 T+p(-1+q_1)q_1 ^p x \beta \right]}
\label{normalizations2}
\end{equation}
and 
\begin{equation}
\overline{\langle s \rangle h}= \overline{  \frac{h^2}{ \left[\lambda -\frac{p}{2T} (1-q^{p-1})\right]^2} } =\frac{1}{Z} \int dh \; P(h) \frac{h^2 (1-q_1) }{T } =-\frac{p q_1^p (-1+q_1)}{2q_1 T+pq_1 ^p(-1+q_1) x \beta }
\ . 
\end{equation}
Then we use the spherical normalization condition, i.e. $\overline{\langle s^2 \rangle} =1$, which in terms of Eq. (\ref{normalizations2}) leads to an additional condition for the breaking parameter $x$:
\begin{equation}
x^{*}= \frac{q_1^{-1-p}\left[ p q_1^p -2 p q_1^{1+p} + p q_1^{2+p} -2q_1^2 T^2\right]} {p(-1+q_1)} \ .
\label{value_x}
\end{equation}
Coming back to Eq. (\ref{motion_virial_app}) and inserting the obtained expression for $x$, we eventually get:
\begin{equation}
\begin{split}
0=&  -\frac{p}{2 T} (1-q_1^{p-1}) \overline{\langle s^2 \rangle} - \overline{h \langle s \rangle} +\frac{p}{2T} \left[ 1 -q_1^{p}(1-x) \right] \Rightarrow \\
0=& -\frac{p}{2T} (1-q_1^{p-1}) + \left . \frac{p q_1^p(-1+q_1)}{ 2 q_1 T + p q_1^p x \beta (-1+q_1) } \right \vert_{x^{*}}+\frac{p}{2} \left( \frac{1 - q_1^{p}}{T} + \frac{q_1^p}{T_\text{eff}} \right) 
\label{equation_spherical_x}
\end{split}
\end{equation}
\begin{equation}
\frac{p}{2T}(1-q_1^{p-1})= - \frac{q_1 T} {1-q_1} +\frac{p}{2} \left( \frac{1 - q_1^{p}}{T} + \frac{q_1^p}{T_\text{eff}} \right)  \ .
\end{equation}
To conclude this part of the computation we can resort to Eq. (\ref{Lagrange_mult_app}) and re-express everything in terms of $\lambda$ as:
\begin{equation}
 \begin{boxed}{\lambda=T+\frac{p}{2} \left( \frac{1-q_1^{p}}{T}+\frac{q_1^{p}}{T_\text{eff}} \right) \ .
}
\label{lagrange_app}
\end{boxed}
\end{equation}

\vspace{0.35cm}

\subsection{3rd approach}

We present now a third, alternative way based on integrating by parts the average spin values, which can be eventually re-expressed in terms of single free-energy differentiation contributions.
Taking advantage of the simplifications performed up to Eq. (\ref{motion_virial_app}), we can directly use the resulting equation
\begin{equation}
\begin{split}
& -T=-\left[ T+\frac{p}{2T}(1-q_1^{p-1}) \overline{\langle s^2 \rangle} + \overline{h \langle s \rangle} \right] +\frac{p}{2 T} \left[1 -q_1^{p}(1-x) \right] \\
& \Rightarrow 0= -\frac{p}{2 T} (1-q_1^{p-1}) \overline{\langle s^2 \rangle} - \overline{h \langle s \rangle} +\frac{p}{2T} \left[ 1 -q_1^{p}(1-x) \right] \ .
\label{motion_app3}
\end{split} 
\end{equation}
Given a generic function $F(h)$, we can write the following expectation value as
\begin{equation}
\begin{split}
h \langle s \rangle_h  = & -\int dh \; P(h)\frac{\partial F}{\partial h} h =-\int dh \; e^{-\frac{h^2}{2 \left( \frac{p}{2} q_1^{p-1} \right) } -\beta x F(h)} \frac{\partial F}{ \partial h} h= \\
= & -\left( \frac{p}{2} q_1^{p-1} \right) \int dh \; e^{-\frac{h^2}{2 \left( \frac{p}{2} q_1^{p-1} \right)}} \left( e^{-\beta x F(h)} \frac{ \partial F}{\partial h} \right)^{'} =\\
=& -\left( \frac{p}{2} q_1^{p-1} \right) \int dh \; e^{-\frac{h^2}{2 \left( \frac{p}{2} q_1^{p-1} \right)}} \left[ e^{-\beta x F(h)} (-\beta x) \left( \frac{\partial F} {\partial h} \right)^2 + e^{-\beta x F(h)} \frac{ \partial^2 F}{\partial h^2} \right] \ .
\end{split}
\end{equation}
In this way, the he equation of motion (\ref{motion_app3}) becomes
\begin{equation}
\begin{split}
\frac{p}{2T}(1-q_1^{p-1}) = &\frac{p}{2} q_1^{p-1} \int dh \; e^{-\frac{h^2}{2 \left( \frac{p}{2} q_1^{p-1} \right)}}  e^{-\beta x F(h)} (-\beta x) \left ( \frac{\partial F}{\partial h} \right)^2 + \frac{p}{2} q_1^{p-1} \int dh \;  e^{-\frac{h^2}{2 \left( \frac{p}{2} q_1^{p-1} \right)}}  e^{-\beta x F(h)} \left( \frac{\partial^2 F}{\partial h^2 }\right)+ \\
+ & \frac{p}{2 T} \left[ 1 -q_1^p(1-x) \right] 
\end{split}
\label{final_3}
\end{equation}
where we have always used $\overline{\langle s^2 \rangle}=1$.
Specialising the analysis to the spherical model with the following free-energy
\begin{equation}
\begin{split}
F(h)= -T \ln \int ds \; e^{-\frac{1}{T}\left[ \frac{1}{2} \left( \lambda-\frac{p}{2T}(1-q_1^{p-1}) \right)s^2 -h s \right] }=  -T \ln \left[ \mathcal{N} e^{\frac{h^2}{ 2 T \left[ \lambda-\frac{p}{2T}(1-q_1^{p-1}) \right] }} \right] 
\end{split}
\end{equation}
whose normalization factor is
\begin{equation}
    \mathcal{N} = \frac{ \sqrt{ 2\pi}}{\sqrt{\frac{1}{T} \left[ \lambda- \frac{p}{2T}(1-q_1^{p-1})  \right]  } } \ ,
    \end{equation}
we have then proposed another way to derive the expression of the spherical parameter $\lambda$.
\vspace{0.3cm}

\hrulefill 

\vspace{0.2cm}

\section{Ising spin model: failure of the previously proposed approach}

For a discrete model we can in principle apply the same procedure starting from the usual equation of motion
\begin{equation}
\biggl \langle s \frac{ds}{dt} \biggr \rangle=-\biggl \langle s \frac{ \partial V}{\partial s} \biggr \rangle + \frac{p(p-1)}{2} \int_{0}^{t} dt' \; R(t,t') C^{p-2}(t,t')\langle s(t)s(t') \rangle   +\langle \xi(t) s(t) \rangle \ ,
\end{equation}
from which, by integrating by parts the first term on the RHS, we eventually obtain:
\begin{equation}
-T=-\frac{1}{Z} \Biggl \lbrace \int ds \; \left[ s(-T) \left ( \frac{ \partial}{\partial s} e^{-V(s)/T} \right) e^{\frac{p}{4T^2}(1-q^{p-1})s^2 +\frac{hs}{T}} \right] \Biggr \rbrace + \frac{p^2}{2} \int_0^{t} R(t,t') C^{p-1}(t,t') dt' \ .
\label{motion_Isingspins}
\end{equation}
At this level, we only need to determine the first two expectation values of the spins according to a $1$RSB Ansatz:
\begin{equation}
0=-\frac{p}{2T}(1-q_1^{p-1})\overline{\langle s^2 \rangle} -\overline{h \langle s \rangle} +\frac{p^2}{2} \int_0^t \frac{x}{T} \frac{\partial C}{\partial t'}(t,t') C^{p-1}(t,t') dt' \ .
\label{motion_simplified}
\end{equation}
In the Ising spin case the conditional probability distribution $P(s \vert h)$ reads
\begin{equation}
P(s \vert h)= \frac{e^{\frac {h   s}{T}}}{2 \cosh(h/T)} \ ,
\end{equation}
which is crucial to determine the only non-trivial expectation value
\begin{equation}
\overline{\langle s \rangle^2}= q_1 \equiv \overline{\left(\frac{e^{h/T}-e^{-h/T}}{2\cosh(h/T)}\right)^2}=\overline{\tanh^2(h/T)}
\end{equation}
as the other one is automatically known, $\overline {\langle s^2 \rangle}=1$.
The field distribution in the Ising case is also known, defined throughout the parameter $x \le 1$:
\begin{equation}
P(h) = \frac{1}{Z} e^{-\frac{h^2}{2\left(\frac{p}{2} q_1^{p-1} \right)}} \left(2 \cosh(h/T) \right)^x \ ,
\end{equation}
which allows us to rewrite the equation of motion (\ref{motion_Isingspins}) in the following form
\begin{equation}
0=-\frac{p}{2T} (1-q_1^{p-1}) -\int \frac{dh}{Z} \; e^{-\frac{h^2}{2 \left(\frac{p}{2} q_1^{p-1} \right)}} \left (2 \cosh(h/T) \right)^x \tanh(h/T) h +\frac{p^2}{2} \int \frac{x}{T} \frac{\partial C}{\partial t'}(t,t') C^{p-1}(t,t') dt' \ .
\end{equation}
By integrating the second term on the RHS by parts and distinguishing the equilibrium and the off-equilibrium contributions of the response function, we end up with
\begin{equation}
\begin{split}
0=& -\frac{p}{2T} (1-q_1^{p-1}) -\frac{p}{2} q_1^{p-1} \Biggl \lbrace  \int \frac{dh}{Z} \; e^{-\frac{h^2}{2 \left(\frac{p}{2} q_1^{p-1} \right)} } \left( 2 \cosh(h/T) \right)^x \left(1-\tanh^2(h/T) \right) \frac{1}{T} +\\
+ &  \int \frac{dh}{Z} \; e^{-\frac{h^2}{2 \left(\frac{p}{2} q_1^{p-1} \right)} } \left( 2 \cosh (h/T)\right)^x  \left(\frac{x}{T} \right) \tanh^2( h/T) \Biggr \rbrace  +\frac{p}{2 T} \left[1-q_1^p (1-x) \right]
 \end{split}
\end{equation}
in other words
\begin{equation}
\begin{split}
0=& -\frac{p}{2T} (1-q_1^{p-1}) -\frac{p}{2} q_1^{p-1} \left(  1-(1-x)\overline{\tanh^2(h/T)}\right)  +\frac{p}{2 T} \left[1-q_1^p (1-x) \right]
 \end{split}
\end{equation}
which however leads to a trivial condition.
\vspace{0.4cm}

\textbf{The case $p=2$. -}. We can consider two simple limiting case, for $x=0$ and $x=1$.  
By setting $x=0$ and $p=2$ in the above equation, we simply recover an identity relationship for the overlap $q_1$, that is:
\begin{equation}
0=\frac{1}{T} \Biggl \lbrace -(1-q_1) -  \; q_1 \frac{1}{Z}\int dh \; e^{-h^2/(2 q)} \left[1- \tanh^2(h/T) \right]+(1-q_1^2) \Biggr \rbrace
\end{equation}
where the normalization factor $Z= \sqrt{ 2 \pi q_1}$.
Therefore, we immediately find $q_1= \int d h P(h, x=0) \tanh^2 (h/T)$. 
Furthermore, if we consider the other straightforward limit for $x=1$ -- which essentially corresponds to an expansion around the plateau, as performed in \cite{Parisi2013} -- we find an automatically satisfied relation for the field distribution $P(h)$, namely $\int dh \; P(h) =1$.

%\CC{I would not say this is surprising. For Ising we need one equation less. Therefore the fact that the equation that in the spherical models leads to a condition on the spherical parameter is trivial (in all its limits) is not very surprising. I would cut this sentence unless there is something I did not understand in your comment about [73].}

\subsection{Double-well potential: perturbative expansion in the limit of a infinitely narrow double well}

To better investigate the peculiarities of the different models, we have also considered a double-well potential 
\begin{equation}
    V(s)=\alpha(s^2-1)^2
    \end{equation}
where $\alpha$ is a tunable parameter that modulates the roughness of the given potential.
In the limit of large $\alpha$, we can safely consider the saddle-point approximation and rewrite the potential as a function of the two different symmetric contributions, \textit{i.e.} $P(s,h)= P(+1,h)+ P(-1,h)$. 
We thus perform a harmonic expansion around each minimum obtaining to the leading order
\begin{equation}
\tilde{V}(s)= 4 \alpha (s-1)^2+o\left((s-1)^2\right)
\label{division}
\end{equation}
and similarly for the other minimum, each being of $O(1/\alpha)$. 
By neglecting the contribution of higher-order terms, the two normalization factors accounting respectively for the expansion around $s=1$ and $s=-1$ turn out to be respectively:
\begin{equation}
Z_{1} \propto \int_{-\infty}^{\infty} du \; e^{\left[-\frac{4 \alpha}{T}+\frac{p}{4T^2}(q_d^{p-1}-q_1^{p-1})\right] u^2 +\left[\frac{p}{2 T^2}(q_d^{p-1}-q_1^{p-1}) +\frac{h}{T}\right]u  +\frac{h}{T}+\frac{p}{4 T^2}(q_d^{p-1}-q_1^{p-1})}  
%\\\frac{2 \sqrt{\pi} \; e^{\frac{h^2 q_1 q_d T+4 \alpha p (q_1^p q_d  - q_1 q_d^p)- 16 h q_1 q_d T \alpha}{T(-p q_1^p q_d +p q_1 q_d^p+16 q_1 q_d T \alpha)}}  }{\sqrt{\frac{-p q_1^p q_d + p q_1 q_d^p+16 q_1 q_d T \alpha}{q_1 q_d T^2}}} 
\end{equation}
and 
\begin{equation}
Z_{-1} \propto \int_{-\infty}^{\infty} du \; e^{\left[-\frac{4 \alpha}{T}+\frac{p}{4T^2}(q_d^{p-1}-q_1^{p-1})\right] u^2 +\left[-\frac{p}{2 T^2}(q_d^{p-1}-q_1^{p-1}) +\frac{h}{T}\right]u  -\frac{h}{T}+\frac{p}{4 T^2}(q_d^{p-1}-q_1^{p-1})}   \ ,
%\frac{2 \sqrt{\pi} \; e^{\frac{h^2 q_1 q_d T+4 \alpha p (q_1^p q_d  - q_1 q_d^p)+ 16 h q_1 q_d T \alpha}{T(-p q_1^p q_d +p q_1 q_d^p+16 q_1 q_d T \alpha)}}  }{\sqrt{\frac{-p q_1^p q_d + p q_1 q_d^p+16 q_1 q_d T \alpha}{q_1 q_d T^2}}}  \ .
\end{equation}
where $u$ has been introduced to denote the change of variable, {\it i.e.} $u=s-1$ in the first $Z$-contribution and $u=s+1$ in the second one.

The boundary term $\frac{p}{4 T^2} (q_d^{p-1}-q_1^{p-1})$ and the external field $h$ contribute only to tilting the potential, hence favouring the positive (or negative) minimum depending on the relative decrease of the free energy.
Therefore, the resulting expression of the free energy can be written as a sum of the two harmonic contributions:
\begin{equation}
\begin{split}
F(h)= & - T \ln \Biggl \lbrace  \int ds \; e^{-\frac{1}{T} \left[ V(s_{+}) -\frac{p}{4T}(q_d^{p-1}-q_1^{p-1})s_{+}^2 -h s_{+} \right] }  
+  \int ds \; e^{-\frac{1}{T} \left[ V(s_{-}) -\frac{p}{4T}(q_d^{p-1}-q_1^{p-1})s_{-}^2 -h s_{-} \right] }  \Biggr \rbrace\\
& = -T \ln  \left(  \frac{ e^{\frac{h^2 q_1 q_d T+4 \alpha p q_1^p q_d  - 4\alpha  p q_1 q_d^p- 16 h q_1 q_d T \alpha}{-p q_1^p q_d T+p q_1 q_d^pT+16 q_1 q_d T^2 \alpha}}  }{\sqrt{\frac{-p q_1^p q_d + p q_1 q_d^p+16 q_1 q_d T \alpha}{q_1 q_d T^2}}}+ \frac{ e^{\frac{h^2 q_1 q_d T+4 \alpha p \left(q_1^p q_d  - q_1 q_d^p \right) + 16 h q_1 q_d T \alpha}{-p q_1^p q_d T+p q_1 q_d^pT+16 q_1 q_d T^2 \alpha}}  }{\sqrt{\frac{-p q_1^p q_d + p q_1 q_d^p+16 q_1 q_d T \alpha}{q_1 q_d T^2}}} \right) 
\end{split}
\end{equation}
from which, focusing on the $h$-dependent terms and neglecting irrelevant prefactors in the logarithm, we recover in the limit $\alpha \rightarrow \infty$ the well-known relationship for the Ising model, i.e. $F(h)=-T \ln \left( 2 \cosh( h/T) \right)$.

% \frac{1}{Z} e^{ -\frac{h^2}{2 \left( \frac{p}{2} q_1^{p-1} \right) } } \left(  \frac{2 \sqrt{\pi} \; e^{\frac{h^2 q_1 q_d T+4 \alpha p q_1^p q_d  - 4\alpha  p q_1 q_d^p- 16 h q_1 q_d T \alpha}{-p q_1^p q_d T+p q_1 q_d^pT+16 q_1 q_d T^2 \alpha}}  }{\sqrt{\frac{-p q_1^p q_d + p q_1 q_d^p+16 q_1 q_d T \alpha}{q_1 q_d T^2}}}+ \frac{2 \sqrt{\pi} \; e^{\frac{h^2 q_1 q_d T+4 \alpha p \left(q_1^p q_d  - q_1 q_d^p \right) + 16 h q_1 q_d T \alpha}{-p q_1^p q_d T+p q_1 q_d^pT+16 q_1 q_d T^2 \alpha}}  }{\sqrt{\frac{-p q_1^p q_d + p q_1 q_d^p+16 q_1 q_d T \alpha}{q_1 q_d T^2}}} \right)^x \ .

To enter into the details of the computation, we consider as usual the effective equation of motion
\begin{equation}
\biggl \langle s \frac{ds}{dt} \biggr \rangle=-\biggl \langle s \frac{ \partial V}{\partial s} \biggr \rangle + \frac{p(p-1)}{2} \int_{0}^{t} dt' \; R(t,t') C^{p-2}(t,t')\langle s(t)s(t') \rangle   +\langle \xi(t) s(t) \rangle 
\label{eq_Ising}
\end{equation}
which, as long discussed before, can be simplified by integrating by parts. It eventually leads to
\begin{equation}
0=-\frac{p}{2 T} \left( q_d^{p-1} -q_1^{p-1} \right) \overline{\langle s^2 \rangle} -\overline{h \langle s \rangle} + \frac{p}{2 T} \left[ q_d^p -q_1^p (1-x) \right] \ .
\label{motion_eq_app}
\end{equation}
which requires the computation of the following expectation value
\begin{equation}
\begin{split}
\overline{h \langle s \rangle_h } = & -\int dh \; P(h)\frac{\partial F}{\partial h} h =-\int dh \; e^{-\frac{h^2}{2 \left( \frac{p}{2} q^{p-1} \right) } -\beta x F(h)} \frac{\partial F}{ \partial h} h= \\
%= & -\left( \frac{p}{2} q^{p-1} \right) \int dh \; e^{-\frac{h^2}{2 \left( \frac{p}{2} q^{p-1} \right)}} \left( e^{-\beta x F(h)} \frac{ \partial F}{\partial h} \right)^{'} =\\
%=& -\left( \frac{p}{2} q^{p-1} \right) \int dh \; e^{-\frac{h^2}{2 \left( \frac{p}{2} q^{p-1} \right)}} \left[ e^{-\beta x F(h)} (-\beta x) \left( \frac{\partial F} {\partial h} \right)^2 + e^{-\beta x F(h)} \frac{ \partial^2 F}{\partial h^2} \right] =\\
=& -\left( \frac{p}{2} q^{p-1} \right) \int dh \; e^{-\frac{h^2}{2 \left( \frac{p}{2} q^{p-1} \right)}} e^{-\beta x F(h)} \left[  (-\beta x) \left( \frac{\partial F} {\partial h} \right)^2 +  \frac{ \partial^2 F}{\partial h^2} \right] \ .
\end{split}
\label{derivatives}
\end{equation}
and therefore of the first two derivatives of $F(h)$ w.r.t $h$ to be conveniently expanded in powers of $1/\alpha$.
According to Eq. (\ref{derivatives}), the two contributions respectively imply:
\begin{equation}
\left( \frac{\partial F}{\partial h} \right)^2= \Biggl \lbrace -\frac{2 q_1 q_d T \left[h+8\alpha \tanh \left(\frac{16 h q_1 q_d \alpha}{-p q_1^p q_d+p q_1 q_d^p+16 q_1 q_d T \alpha} \right) \right]}{ \left( -p q_1^p q_d + p q_1 q_d^p +16 q_1 q_d T \alpha \right)}\Biggr \rbrace ^2 \ , 
\label{F_prime}
\end{equation} 
\\
\begin{equation}
\frac{\partial^2 F}{\partial h^2}=-\frac{2 q_1 q_d T \left[ -p q_1^p q_d + p q_1 q_d^p+16q_1 q_d T \alpha +128 q_1 q_d \alpha^2 \; \text{sech}^2 \left( \frac{16 h q_1 q_d \alpha}{-p q_1^p q_d + p q_1 q_d^p+16 q_1 q_d T \alpha}\right) \right]}{\left( -p q_1^p q_d + p q_1 q_d^p +16 q_1 q_d T \alpha \right)^2} \ ,
\label{F_second}
\end{equation}
that, once they are expanded to the leading order in $1/\alpha$, yield: 
\begin{equation}
\left( \frac{\partial F}{\partial h} \right)^2 \simeq \tanh^2(h/T) +\frac{1}{ 8 q_1 q_d \alpha} \left[ \tanh( h/T) \left( 2 h q_1 q_d -p(-q_1^p q_d+q_1 q_d^p ) \frac{1}{T} (h/T \; \text{sech}^2(h/T) + \tanh(h/T) \right) \right] + O \left(\frac{1}{\alpha}\right)^2
\end{equation}
\begin{equation}
\frac{\partial^2 F}{\partial h^2}  \simeq -\frac{1}{T} \text{sech}^2(h/T)+ \frac{1}{8 q_1 q_d \alpha} \left[ -p \left( -q_1^p q_d+ q_1 q_d^p \right) \frac{1}{T^2} \text{sech}^2 (h/T) \left(-1+h/T \tanh(h/T)\right) \right] + O \left(\frac{1}{\alpha}\right)^2 
\end{equation}
\vspace{0.5cm} 

$\spadesuit$
In the simplest case, which corresponds to setting the diagonal value $q_d=1$ and the breaking parameter $x=1$, the equation of motion reduces to 
\begin{equation}
\begin{split}
\frac{p}{2 T}  -\frac{p}{2 T} q_1^{p-1}  & = -\frac{p}{2 } q_1^{p-1}\int dh P(h) \; \frac{x}{T} \left(\frac{ \partial F} {\partial h} \right)^2 + \frac{p}{2}q_1^{p-1} \int dh P(h) \frac{\partial^2 F}{\partial h^2} 
+\frac{p}{2 T}  \ ,\\
0 & = 1- \int dh P(h) \left(\frac{\partial F}{\partial h} \right)^2+ \int dh P(h) \frac{\partial^2 F}{\partial h^2} 
\end{split}
\end{equation}
and focusing only on the leading order terms, we would get:
\begin{equation}
0=1- \int dh P(h) \tanh^2(h/T) -\int dh P(h) \text{sech}^2(h/T) \ .
\end{equation}
Again, this equation results into an identity condition for the probability distribution $P(h)$. 
Going further in the expansion of Eqs. (\ref{F_prime})-(\ref{F_second}) and including also higher-order terms in the computation
\begin{equation}
\begin{split}
0= & - \frac{p}{2T} q_1^{p-1} \int dh P(h) \left[ \frac{1}{ 8 \alpha} \tanh(h/T) 2 h+ \frac{ p(q_1^p-q_1)}{8 q_1 T^2 \alpha} h \tanh(h/T) \text{sech}^2 (h/T) + \frac{ p (q_1^p-q_1)}{8 q_1 T \alpha} \tanh^2(h/T)\right]+\\
& - \frac{p}{2 T} q_1^{p-1} \int dh P(h) \left[\frac{ p(q_1^p-q_1)}{8 q_1  T \alpha} \left(\text{sech}^2(h/T ) - \frac{1}{T} h \tanh(h/T) \text{sech}^2(h/T ) \right) \right]+O \left( \frac{1}{\alpha} \right) 
\end{split}
\end{equation}
we nevertheless notice that the last term -- that might be possibly simplified by integration by parts -- cancels out with the same term of opposite sign in the first line. 

The problem is thus solved in the case of a spherical $p$-spin, but not in more general cases. Even for the corresponding soft-spin version of an Ising model, based on the introduction of a tunable parameter $\alpha$ modulating the roughness of the double-well potential, the equation above appears to be needless and to provide only very basic information. 

From this analysis, we conclude that the virial equation (\ref{eq_Ising}) is nothing more than an equation for the correlation $C(t,t')$ at equal times that, in the case of an Ising model, is automatically satisfied whereas in the spherical model leads to an additional condition useful to fix the spherical parameter. To determine the effective temperature and close the system of equations, we need to define an additional condition to be mapped on the analog of an equation for the response function. 

\newpage

\section{Connection between effective temperature and breaking parameter of the static solution}

In the case of the spherical $p$-spin model, we have all the tools to show the underlying mapping between the effective temperature $T_\text{eff}$ and the breaking point $x$ of the static solution.
The stationary field distribution has a simple quadratic dependence on the effective field and can be easily manipulated to get all other missing information.
\begin{equation}
\begin{split}
P(h)= & \exp \Biggl \lbrace -\frac{h^2}{T_\text{eff}} \left[ \frac{1}{ \frac{2 p q_1^{p-1}}{2 T_\text{eff}}} -\frac{1}{ 2\left( \lambda - \frac{p}{2 T} (1-q_1^{p-1}) \right) }  \right] \Biggr \rbrace=\\ \\
&= \exp \Biggl \lbrace -\frac{h^2}{ T_\text{eff}} \; \frac{\lambda- \frac{p}{2 T}(1-q_1^{p-1})-\frac{p}{2 T_\text{eff}} q_1^{p-1}}{2\frac{p}{ 2 T_\text{eff}} q_1^{p-1} \left( \lambda-\frac{p}{2 T}(1-q_1^{p-1}) \right) } \Biggr \rbrace 
\end{split}
\end{equation}

In Secs. (\ref{aging_1RSB}) and (\ref{applications}) we have shown that
\begin{equation}
\begin{split}
\overline{\langle s \rangle}= & \frac{\int_{-\infty}^{\infty} P(s \vert h) s }{\int_{-\infty}^{\infty} P(s \vert h)}=\frac{1}{\lambda-\frac{p}{2T}(1-q_1^{p-1})} \left[ \frac{p(p-1)}{2} \int_0^{t} R_A(t,t') C_A(t,t')^{p-2} s(t') dt'+ \xi_A \right] 
=\\
= & \overline{ \frac{h}{ \lambda -\frac{p}{2T} (1-q^{p-1})} } \ .
\end{split}
\end{equation}
to be eventually averaged over the effective field distribution $P(h)$.
Using the information on the average spin value, we can rewrite the equation over the off-diagonal value of the overlap matrix, $q_1$, as 
\begin{equation}
q_1= \frac{\frac{p}{2} q_1^{p-1} \left[\lambda-\frac{p}{2 T} (1-q_1^{p-1}) \right] }{\left[ \lambda-\frac{p}{2 T} (1-q_1^{p-1}) \right]^2\left[\lambda- \frac{p}{2 T}(1-q_1^{p-1})-\frac{p}{2 T_\text{eff}} q_1^{p-1}\right]}
\end{equation}
and by simple algebraic manipulations get the following expression:
\begin{equation}
\left[ \lambda- \frac{p}{2 T} (1-q_1^{p-1})\right] \left[ \lambda- \frac{p}{2T}(1-q_1^{p-1}) -\frac{p}{2 T_\text{eff}} q_1^{p-1} \right] = \frac{p}{2} q_1^{p-2} \ .
\label{q1-Teff}
\end{equation}
The first parenthesis can be rewritten in a more straightforward way by using the condition for the Lagrange multiplier, as derived in Eq. (\ref{Lagrange_mult_app}), {\it i.e.} $\lambda-\frac{p}{2T}(1-q_1^{p-1})= T/( 1-q_1)$. The above equation becomes then
\begin{equation}
\frac{T^2}{(1-q_1)^2}-\frac{T}{(1-q_1)} \frac{p}{2T_\text{eff}} q_1^{p-1}=\frac{p}{2}q_1^{p-2} \hspace{0.6cm} \rightarrow \hspace{0.6cm} 1-\frac{(1-q_1)}{T} \frac{p}{2 T_\text{eff}} q_1^{p-1}= \frac{p}{2T^2} q_1^{p-2} (1-q_1)^2 \ .
\end{equation}
To extract a resulting equation for the effective temperature we can recall the condition obtained in the main text in terms marginal stability, which has been imposed on the TTI dynamics.
Then, using Eq. (\ref{replicon_spherical}) and simply equating the RHS to $1/(p-1)$, we obtain:
\begin{equation}
p-2 -\frac{(1-q_1)}{T} \frac{p(p-1)}{2 T_\text{eff}} q_1^{p-1}=0 \ .
\end{equation}
We have thus recovered the relationship between the effective temperature and the breaking parameter $x$ within the $1$RSB approximation in the replica formalism for the spherical $p$-spin model:
\begin{equation}
x \equiv \frac{T}{T_\text{eff}}=\frac{(p-2)(1-q_1)}{q_1} \ .
\end{equation}
The resulting value of the breaking parameter $x$ corresponds to those TAP states which are marginally stable, the so-called \emph{threshold states}. The critical slowing down of the dynamics and related aging phenomena are then consequences of the flatness of the free energy around these states.

\newpage

\section{Connection with previous formalisms and identification of the \emph{anomaly}}
\label{RM}

%Let us consider now a manifold embedded in a random medium, which can described by a displacement field and the following Hamiltonian \cite{Franz1994, Cugliandolo1996RM}:
%\begin{equation}H= \int d x \left[ \frac{c}{2} \left( \nabla {\phi}(x) %\right)^2 + V(\phi(x),x) + \frac{\lambda}{2} \phi^2 %\right]
%\end{equation} %\mu$ being a mass term which constrains the manifold to fluctuate in a given volume, and $V$ a Gaussian random potential with correlation:
%\begin{equation} \overline{V(\phi, x) V(\phi^{'}, x^{'})}= -N\delta %(x-x') \mathcal{V} \left( \frac{(\phi-\phi')^2}{N} %\right)\end{equation}
%Then, for a $D=0$ manifold, the correlation and response functions become:\begin{equation}C(t,t')= \frac{1}{N} \overline{ \langle \phi(t) \phi(t') \rangle } \hspace{0.4cm} \ , \hspace{0.7cm} R(t,t')= \frac{1}{N} \left . \overline{ \frac{\delta  \langle \phi(t) \rangle}{ \delta h(t')}} \right \vert_{h=0} \ .
%\end{equation}

To prove the extreme generality of our approach we have also considered the problem of particle in a random manifold that has been extensively studied in the past \cite{Franz1994, Cugliandolo1996RM} and from which, under suitable assumptions, the usual equations for the spherical $p$-spin model can be recovered.
The mean-field dynamical equations the two-time correlation and response functions can be expressed in the following form:
\begin{equation}
\frac{ \partial C(t,t')}{\partial t}= - \lambda C(t,t') + \frac{p}{2} \int_{0}^{t'} d s \;  C^{p-1}(t,s) R(t',s) - \frac{p(p-1)}{2} \int_{0}^{t} d s \; C^{p-2}(t,s) R(t,s) \left[ C(t,t')-C(s,t') \right]+ 2 T R(t',t) \ ,
\label{correlation_part}
\end{equation}
\begin{equation}
\frac{ \partial R(t,t')}{\partial t}= - \lambda R(t,t') -\frac{p(p-1)}{2} \int_{0}^{t} d s \; C^{p-2}(t,s) R(t,s) \left[ R(t,t')-R(s,t') \right]
\label{response}
\end{equation}
where the function must satisfy the following prescriptions 
according to the causality property and the Ito integration scheme:
\begin{equation}
R(t,t)=0 \hspace{0.05cm} \ , \hspace{0.4cm} 
\lim_{\epsilon \rightarrow 0} R(t,t-\epsilon)=1 \hspace{0.05cm} , 
\hspace{0.4cm} R(t', t)=0 \hspace{0.4cm} \text{if} \hspace{0.4cm} t>t'
\end{equation}
Therefore, the last term in the equation for the correlation function vanishes and, as $t' \rightarrow t$, Eq. (\ref{correlation_part}) becomes
\begin{equation}
\frac{1}{2} \frac{\text{d} C(t,t)}{ \text{d}t}= -\lambda C(t,t)+ T + \frac{p}{2} \int_{0}^{t} d s \; C^{p-1} (t,s) R(t,s)+ \frac{p(p-1)}{2} \int_{0}^{t} d s \; C^{p-2}(t,s) R(t,s) \left[ C(t,t)-C(s,s)+ B(t,s) \right]
\end{equation}
or, equivalently, in terms of the mean-squared displacement $B(t,t')$, which is defined as 
\begin{equation}
B(t,t') \equiv C(t,t)+ C(t',t')- 2 C(t,t')= \overline{ \langle \left[ s(t)-s(t') \right]^2 \rangle } \ .
\end{equation}

If the correlation is set to $q$, the above equation for the total correlation evaluated for $t \approx t'$ reduces to
\begin{equation}
\lambda(t) q = T +  \int_{0}^{t} d s \left[ \frac{p}{2} C^{p-1} (t,s)+\frac{p(p-1)}{2} C^{p-2}(t,s) B(t,s) \right] R(t,s) \ .
\end{equation}
The Lagrange multiplier has an explicit dependence on time and has to be properly fixed accordingly to the condition on the spherical constraint. 
If we impose $q=1$ and simplify the product of the different combinations of correlations, we recover exactly the same expression as in (\ref{lagrange}), which has been obtained before in Eq. (\ref{lagrange_app}) by using a virial expansion, namely:
\begin{equation}
\begin{boxed}{
\lambda= T+ \frac{p^2}{2} \int_0^{t} R(t,t') C^{p-1}(t,t') dt'}
\end{boxed}
\end{equation}

\subsection{Additional equation on the response function and derivation of the marginal stability condition}

We want now to consider the second equation, for the response function, and thus derive the analogue of the \emph{anomaly}, which accounts for those times that are not included in the asymptotic regime but for which aging effects are nevertheless relevant. The anomaly essentially couples the asymptotic time regime, for which $t-t'$ is finite, with the non-asymptotic dynamical contribution.
%By using Eq. (\ref{response}), one can eventually write \cite{Cugliandolo1996RM}:
%\begin{equation}
%0=r(t,t') \left[ - \lambda + 4 \int_{0}^{t} ds \mathcal{V}''(b(t,s)) r(t,s) - \frac{2 q}{T} \mathcal{V}''(q) \right]\label{random_manifold_response}
%\end{equation}
%where the function $b(t,t)$ can take the following values:
%\begin{equation}
%b(t,t)=0 \hspace{0.01cm} \ , \hspace{0.2cm} %b(t,t_{-})= q \hspace{0.01cm} \ , \hspace{0.2cm} \lim_{\tau \rightarrow \infty} b(\tau)=q \hspace{0.01cm} \ , \hspace{0.2cm} b(t,0_{+})= b_0 \ .
%\end{equation}
The anomaly is zero in the high-temperature phase and takes a non-vanishing contribution in the aging regime, associated with a finite value of the overlap parameter \cite{Franz1994}.

The equation for the response function analysed in the aging regime implies
\begin{equation}
0= \left[ -\lambda_{\infty}+ \frac{p}{2 T} \left(1-q_{1}^{p-1} \right) \right] R_{\text{A}}(t,t') + \frac{p(p-1)}{2} R_{\text{A}}(t,t') C_{\text{A}}^{p-2}(t,t')  \frac{(1-q_1)}{T}+ \frac{p(p-1)}{2} \int_{t'}^{t} {R_\text{A}}(s,t'') C_{\text{A}}^{p-2}(s,t'') R_{\text{A}}(t'',t') d s
\end{equation}
which, for $t' \approx t$, becomes:
\begin{equation}
0= R_\text{A}(t,t') \left[ - \lambda_{\infty} + \frac{p}{2T} (1-q_1^{p-1}) + \frac{p(p-1)}{2}C_\text{A}^{p-2}(t,t') \frac{1-q_1}{T}\right] \ .
\end{equation}
By plugging the asymptotic value of the Lagrange multiplier in the previous expression
\begin{equation}
\lambda_{\infty}= \frac{T}{(1-q_1)}+ \frac{p}{2 T} (1-q_1^{p-1}) \ ,
\end{equation}
we immediately get
\begin{equation}
0=R_\text{A}(t,t') \left[ -\frac{T}{(1-q_1)}- \frac{ p}{2 T} (1-q_1^{p-1})+ \frac{p }{2 T} (1-q_1^{p-1}) + \frac{{p (p-1)}}{2 T} q_1^{p-2} (1-q_1) \right]
\end{equation}
By requiring that the response function is nonzero, we can obtain the marginality condition in an alternative way
\begin{equation}
\begin{boxed}{
\frac{1}{p-1}= \left(\frac{ p}{2 T^2}\right) q_1^{p-2}(1-q_1)^2  }
\end{boxed}
\end{equation}
which precisely corresponds to the appearance of a vanishing replicon eigenvalue in the stability matrix in a static replica formalism.

\vspace{0.6cm}

\bibliography{dynam}

\end{document}